\documentclass[table,pdflatex,sn-mathphys]{sn-jnl}
\usepackage{amsmath, amssymb,amsfonts}
\usepackage{bm}
\usepackage{graphicx}
\usepackage{color}
\usepackage{mathrsfs}
\usepackage{soul,ulem}
\usepackage{natbib}
\usepackage{caption}
\usepackage{subcaption}
\usepackage[pagewise]{lineno}
\newcommand{\abhi}[1]{{\color{black} #1}}

\def\bSig\mathbf{\Sigma}

\newcommand{\bs}[1]{\mathbf{#1}}
\newcommand{\bss}[1]{\boldsymbol{#1}}
\newcommand{\shade}[1]{{\cellcolor[HTML]{C0C0C0}#1}}
\newtheorem{remark}{Remark}%
\begin{document}
\title[Joint modeling of cyclical longitudinal process and discrete survival time]{Joint modeling of geometric features of longitudinal process and discrete survival time measured on nested timescales: an application to fecundity studies}

\author[1]{\fnm{Abhisek } \sur{Saha}}

\author[2]{\fnm{Ling } \sur{Ma}}

\author[3]{\fnm{Animikh } \sur{Biswas}}

\author*[1]{\fnm{Rajeshwari} \sur{Sundaram}}\email{sundaramr2@mail.nih.gov}

\affil*[1]{\orgdiv{{\it Eunice Kennedy Shriver} National Institute of Child Health and Human Development}, \orgname{ NIH}, \orgaddress{\street{DHHS,6710B Rockledge Dr. }, \city{Bethesda}, \postcode{20817}, \state{Maryland}, \country{U.S.A}}}

\affil[2]{\orgdiv{Principal Statistical Scientist}, \orgname{Genentech Inc}, \orgaddress{ \city{San Francisco},  \state{California}, \country{U.S.A}}}

\affil[3]{\orgdiv{Department of Mathematics and Statistics}, \orgname{University of Maryland Baltimore County}, \orgaddress{  \state{Maryland}, \country{U.S.A}}}

\abstract{ In biomedical studies, longitudinal processes are collected till time-to-event, sometimes on nested timescales  (example, days within months). Most of the literature in joint modeling of longitudinal and time-to-event data has focused on modeling the mean or dispersion of the longitudinal process with the hazard for time-to-event. However, based on the motivating studies,  it may be of interest to investigate how the cycle-level {\it geometric features} (such as the curvature, location and height of a peak), of a cyclical longitudinal process is associated with the time-to-event being studied. We propose a shared parameter joint model for a cyclical longitudinal process and a discrete survival time, measured on nested timescales, where the cycle-varying geometric feature is modeled through a linear mixed effects model and a proportional hazards model for the discrete survival time. The proposed approach allows for prediction of survival probabilities for future subjects based on their available longitudinal measurements. Our proposed model and approach is illustrated through simulation and analysis of Stress and Time-to-Pregnancy, a component of Oxford Conception Study. A joint modeling approach was used to assess whether the cycle-specific geometric features of the lutenizing hormone measurements, such as its peak or its curvature, are associated with time-to-pregnancy (TTP).}



\maketitle
\section{Introduction}\label{sec1}
In biomedical studies, information is routinely collected longitudinally on various  biomarkers up to a time-to-event (usually censored), along with additional covariates. An often cited example of such data is from HIV clinical trial with the key longitudinal process of interest being CD4 counts. Many other examples from cancer studies, reproductive health etc. have been the motivation for the development of various methods in this area of research. In joint modeling, one is typically interested in (i) how to model the pattern of change of the longitudinal process, and (ii) to characterize the relationship between the survival event, the longitudinal process, and the covariates. 

Most of the literature on joint modeling of longitudinal process and time-to-event have focused on modeling the mean of the longitudinal process, where the dependence between the longitudinal process and time-to-event is through the mean process with subject-specific random effect(s) \citep{Wu:Carroll:1988, Tsiatis:Davidian:2004}. Joint modeling in the context of longitudinal process and discrete time-to-events was first studied by Albert et al. (2010) \cite{albert2010approach}, and later by Qiu et al. (2016)\cite{qiu2016joint} in the context of shared parameter framework approach. Recently, some authors have also considered modeling the dispersions of the longitudinal process and time-to-event 
\citep{gao2011joint,Mclain:Lum:Sundaram:2012}. In many situations (the motivating example discussed below), it may be more of interest to study the geometric features of a cyclical longitudinal process. Another common aspect of the existing joint modeling literature is that they focus on the situations where the longitudinal process and the time-to-event are on the same time scale, while there are examples where this assumption may not be true. For instance, in our motivating example, the longitudinal process is measured on a daily scale while the time-to-pregnancy is measured in menstrual cycles (approximately monthly). In this paper, our main objective is to jointly model cyclic longitudinal process with time-to-event, measured on nested timescales, where the dependence between the cyclic longitudinal process and time to event is captured through cycle-level geometric features of the longitudinal process with subject-specific random effects.

We now present our motivating example from reproductive health studies. Reproductive hormones, like the luteinizing hormone (LH), estrogen and its metabolites (eg, estrone-3-gluconoride  (e3g) etc) patterns play an important role in the study of conception, infertility and other chronic disease (\citealp{Mumford:etal:2012,Parazzini:etal:1993,Terry:etal:2005,Whelan:etal:1994,Solomon:etal:2001,Solomon:etal:2002}). 
However, data on these endogenous hormones is difficult to quantify due to complex cyclical patterns of hormones, the need for timed collection, and the cost required for multiple 
sample collections. So, much of research has focused on menstrual cycle characteristics  such as cycle length  as proxies for cumulative hormonal exposure and/or hormonal patterns as they can be easily assessed in population studies.  Short and long or irregular cycles have been associated with increased risks for breast cancers, osteoporosis, type 2 diabetes mellitus, cardiovascular diseases etc. Consequently, it is important to be able to study the hormonal profiles directly. LH has important role in the luteinizing of the follicle and the functional maturation of the nucleus of the oocyte \citep{Verpoest:etal:2000}. Abnormalities in the LH surge may impair the development of the oocyte and consequently it's fertilization ability. Additionally, LH surge abnormalities, such as reduced peak values, have been associated with infertility  \citep{Cohlen:etal:1993,Cahill:etal:1997}, indicating that shape of the hormonal curve plays an important biological role as well. Moreover, the pattern of the LH profile is highly variable in normal menstruating women. Motivated by these issues, we are interested in modeling the LH surge and its relationship with time-to-pregnancy. A source of such data arises in the prospective pregnancy studies, where couples are followed from the time they go off contraception (with the intention of becoming pregnant) until they get pregnant. In these studies, women use ovulation kits which track the LH and e3g daily within a fixed window to precisely capture the impending day of ovulation in each menstrual cycle. This provides an opportunity to study the patterns of surge of these hormones, especially LH and its relationship with time-to-pregnancy. Motivated by scietific hypothesis, we focus on three geometric features, namely, the value of LH peak, the curvature at the LH peak and the average curvature of the LH profile within fertile window (the window of opportunity around ovulation for conception in a menstrual cycle) from the longitudinally measured LH values within each menstrual cycle.

Motivated by the example described above, we consider joint modeling of the cyclic longitudinal process (``hormonal profile") and a discrete survival time (``time-to-pregnancy"), where one is interested in modeling various { \it geometric features} of the longitudinal process at the cycle-level (e.g., value at the peak, curvature at the peak, average curvature within fertile window of a cycle). In the next section, we introduce our data and the modeling framework. In Section 3, we provide the estimation approach for the parameters of interest and also discuss the prediction approach for the time-to-event distribution of a new subject given its longitudinal measurement history. We assess the performance of the proposed estimates through simulation studies in Section 4. In Section 5, we present detailed analysis of the Stress and Time-to-pregnancy, a sub-component of the Oxford Conception Study \citep{louis2011stress, McLain:Sundaram:Louis:2015}.
.
\section{Model and Notation}
\label{s:model}
For woman $i$, $i=1, \dots, n$, let $T_i$ denote the time-to-pregnancy, i.e., the number of menstrual cycles it took to get pregnant. As is typical in time-to-event studies, $T_i$ is subject to right censoring and one observes $X_i=\min(T_i,\tau_i)$ and $ \delta_i=I(T_i\leq\tau_i)$,  where $\tau_i$ denotes the censoring time and $I(\cdot)$ denotes the indicator function. Throughout this article, we assume that the censoring time $\tau_i$ is independent of $T_i$.

In prospective pregnancy studies, ovulation kits are used by each woman participant to identify the day of ovulation. These kits typically require testing to be done on day $6$ through day $25$ of every menstrual cycle for measuring the hormone levels to identify the day of ovulation within each cycle. So, essentially the test results in underlying hormonal data on LH from day $6$ through day $25$ in each cycle for every woman. We denote the hormonal data in a cycle as $h(t)$, $t\in[L, M]$ where $L, M$ are days from the start of cycle where hormone measurements are collected within each menstrual cycle and $\hat{t}$ is the time point from the start of the cycle where the hormonal profile reaches the highest level (peak). Also, we define the fertile window ``the window of opportunity for conception around ovulation" in a cycle as $[\hat{t}-5,\ \hat{t}+1].$ The three geometric features of hormonal profile that are of interest here are: (1) cycle-specific curvature at the hormone peak, $k(\hat{t})=\vert h''(\hat{t})\vert \{1+h'(\hat{t})\}^{-3/2}$, which measures the sharpness of the hormonal profile at the peak time; (2) cycle-specific hormone peak value, $h(\hat{t})$; (3) the average curvature of the hormone profile within fertile window, $\sum_{t=\hat{t}-5}^{\hat{t}+1}k(t)/6$; within each cycle. 

Let $\tilde{Y}_{ij}$ be the true geometric feature of hormonal profile of interest for $i$-th woman in menstrual cycle $j$ and let $Z_{ij}$ be the vector of covariates for $\tilde{Y}_{ij}$. Denote by $Y_{ij}$ the observed geometric feature of hormonal profile for  $i$-th woman in menstrual cycle $j$. The observed geometric features for each cycle are then calculated from the observed hormone data, first by using some smoothing technique, e.g. B-splines and then calculating the feature based on the formula mentioned in the previous paragraph. 

We relate the true geometric feature $\tilde{Y}_{ij}$ to the covariates $Z_{ij}$ through a linear model with subject-specific random intercept $b_{Y,i}$,
$$\tilde{Y}_{ij}=Z_{ij}'\bss{\beta}+b_{Y,i},$$ \abhi{where $\bss{\beta}$ is the vector of regression coefficients corresponding to $Z_{ij}$.} The observed hormonal geometric features $Y_{ij}$ are then modeled by
\begin{equation}
	\label{hormone_model}
	Y_{ij}=\tilde{Y}_{ij} +\epsilon_{ij},
\end{equation}
where $\epsilon_{ij}$ are all independent and identically distributed and follow $N(0,\sigma^2)$ distribution.

We use the discrete survival model by \abhi{Sundaram et al. (2012)} \cite{Sundaram2:McLain:Louis:2012} for TTP where the hazard of discrete survival time is related linearly to the covariates when transformed by a complementary log-log function. It also accounts for the fact that, the hazard for conception in a cycle is zero if the couple does not have any intercourse in the fertile window of that cycle. The model is given by
\begin{equation}
	\label{TTP_sundaram}
	\lambda_i(j \mid b_{T,i},U_{ij})=1-\exp\Big[-A_{ij}\exp\{b_{T,i}+\rho_j+U_{ij}'\bss{\gamma}\}\Big]
\end{equation}
where for subject $i$ in cycle $j$, $U_{ij}$ is the vector of covariates for TTP (which could have overlap with $Z_{ij}$, the covariates for the geometric features), $A_{ij}$ is the indicator of intercourse within fertile window, $\rho_j$ is the baseline effect for cycle $j$, \abhi{$\bss{\gamma}$ represents the regression coefficients of $U_{ij}$} and $b_{T,i}$ is a subject-specific random effect. Recall that the fertile window refers to the days in a menstrual cycle around the day of ovulation when a couple having intercourse can potentially conceive; $A_{ij}=0$ means that couple $i$ did not have intercourse during the fertile window of cycle $j$, which implies that there is no risk of pregnancy in that cycle.

To study the association between a woman's hormonal profile and TTP, we take into account the cycle-level geometric feature of a woman's hormonal profile, $\tilde{Y}_{ij}$, in (\ref{TTP_sundaram}). Recalling that $\tilde{Y}_{ij}=Z_{ij}'\bss{\beta}+b_{Y,i}$, we propose the following model
\begin{eqnarray}
	\label{TTP_model}
	\lambda_i(j \vert b_{T,i}, b_{Y,i}, U_{ij}) &=& 1-\exp\Big[-A_{ij}\exp\{b_{T,i}+\rho_j+U_{ij}'\bss{\gamma}+\psi_\mu\tilde{Y}_{ij}\}\Big] \\
	&=&1-\exp\Big[-A_{ij}\exp\{b_{T,i}+\rho_j+U_{ij}'\bss{\gamma}+\psi_\mu(Z_{ij}'\bss{\beta}+b_{Y,i})\}\Big],\nonumber
\end{eqnarray}
where $\psi_\mu$ is the regression coefficient of $\tilde{Y}_{ij}$,
and assume that $\bs{b}_i \equiv (b_{Y,i}, b_{T,i}) $ follows   \abhi{multivariate normal distribution with mean $\bs{0}$ and variance covariance matrix $D$, \textbf{MVN}$(\bs{0},D)$,} with
$$ D=\left( \begin{array}{cc}
	\sigma_1^2 & \zeta\sigma_1\sigma_2 \\
	\zeta\sigma_1\sigma_2 & \sigma_2^2
\end{array}
\right),$$
and $\bs{b}_i$'s are independent and identically distributed (iid) and independent of $\epsilon_{ij}$. The association between the hormonal profile and TTP is taken into account not only through the fact that the cycle-level geometric feature is included in the model for TTP as a predictor, but also through the possible correlation between subject specific random effects $b_{Y,i}$ and $b_{T_i}$. 

\subsection{Estimation and Prediction}
\label{s:estimation}
Denote the observed data for subject $i$ as $O_i=(X_i,\delta_i,Y_{ij},Z_{ij},U_{ij}, 1 \leq j \leq X_i)$. The observed data log likelihood can be written as
\begin{equation}
	\label{loglike}
	l(\bss{\theta})=\sum_{i=1}^{n}\log\bigg\{\int f_i(X_i\vert\bs{b})^{\delta_i}S_i(X_i\vert\bs{b})^{1-\delta_i}\prod_{j=1}^{X_i}f_{Y_{ij}}(Y_{ij}\vert\bs{b})f_b(\bs{b})d\bs{b}\bigg\},
\end{equation}
where $\bss{\theta}=(\bss{\beta},\bss{\gamma},\bss{\rho},\psi,\sigma_1^2,\sigma_2^2,\zeta)'$, $\bs{b}=(b_Y, b_T)$,
\[ S_i(j\vert\bs{b})=\exp\bigg\{-\sum_{k=1}^{j}A_{ik}\exp(b_T+\rho_k+U_{ik}'\bss{\gamma}+\psi_\mu(Z_{ik}'\bss{\beta}+b_Y))\bigg\},
\]
\[f_i(j\vert\bs{b})=S_i(j-1 \vert\bs{b})-S_i(j\vert\bs{b}),
\]
and
$f_{Y_{ij}}(Y_{ij} \vert \bs{b})$ is the density function of normal distribution with mean $(Z_{ij}'\bss{\beta}+b_Y)$ and variance $\sigma^2$. 

One natural way of finding estimates for $\bss{\theta}$ is to maximize the log likelihood function (\ref{loglike}) with respect to $\bss{\theta}$. However, the two-dimensional integration with respect to the random effects does not have a closed form. We propose to use Gaussian quadrature for approximation. Specifically, let  \abhi{$\bs{b}=\tilde{\bs{Z}}R$} where $R$ is the Choleskey square root of the covariance matrix $D$ (e.g. $D=R'R$) and  \abhi{$\tilde{\bs{Z}}$} is a two-dimensional row-vector of independent standard normal variables. Let $\{(\tilde{Z}_k,w_k), k=1,\dots,K\}$ be the $K$ Gaussian nodes and weights for a standard normal variable, then the $K^2$ nodes of $\bs{b}$ may be constructed by 
$$ \bs{b}_{k,s}=\abhi{(\tilde{Z}_k,\tilde{Z}_s)R}=(R_{11}\tilde{Z}_k+R_{21}\tilde{Z}_s, R_{12}\tilde{Z}_k+R_{22}\tilde{Z}_s), k,s = 1,\dots, K,$$ 
where $R_{ks}$ is the $(k,s)$th element of $R$, and the associated weight is calculated by $w_kw_s$. Then the likelihood contribution of the $i$-th subject could be approximated by
\begin{eqnarray}
	\nonumber&\int f_i(X_i\vert\bs{b})^{\delta_i}S_i(X_i\vert\bs{b})^{1-\delta_i}\prod_{j=1}^{X_i}f_{Y_{ij}}(Y_{ij}\vert\bs{b})f_b(\bs{b})d\bs{b}
	\\
	\nonumber \approx& \sum_{k=1}^{K} \sum_{s=1}^{K}f_i(X_i\vert\bs{b}_{k,s})^{\delta_i}S_i(X_i\vert\bs{b}_{k,s})^{1-\delta_i}\prod_{j=1}^{X_i}f_{Y_{ij}}(Y_{ij}\vert\bs{b}_{k,s})w_kw_s.
\end{eqnarray} 
We then maximize the approximated log likelihood function to get estimate of $\bss{\theta}$, denoted by $\hat{\bss{\theta}}$. The covariance matrix $\Sigma$ of $\bss{\theta}$ is estimated by using the observed information matrix.

One nice feature of using the joint modeling approach is that we could use hormonal profile characteristics to predict TTP distribution. Based on a joint model fitted on a sample of size $n$, we are interested in predicting the time-to-event distribution for a new subject $i$ that has provided a set of longitudinal measurements up to cycle $j_0$. Denote $\hat{\Sigma}=\hat{\textrm{var}}(\hat{\bss{\theta}})$ as the estimated variance covariance matrix for $\bss{\theta}$. The partial information for the new subject $i$ is denoted by $\mathcal{D}_i(j_0)=\{\mathcal{Y}_i(j_0),U_i(j_0),Z_i(j_0)\}$, where $\mathcal{Y}_i(j_0)=\{Y_{ij}, j\leq j_0\}$, $U_i(j_0)=\{U_{ij}, j\leq j_0\}$ and $Z_i(j_0)=\{Z_{ij}, j\leq j_0\}$. Prediction of the conditional probability of surviving cycle $j$ is of interest only if the couple have not achieved pregnancy at cycle $j_0$. Hence we focus on the conditional probability of surviving cycle $j$ given survival up to cycle $j_0$.
\begin{eqnarray}
	\label{cond_surv}
	\nonumber\pi_i(j\mid j_0)&\equiv& \textrm{Pr}(T_i\geq j\mid T_i>j_0,\mathcal{D}_i(j_0),\mathcal{D}_n)\\
	&=& \int \textrm{Pr}(T_i\geq j\mid T_i> j_0, \mathcal{D}_i(j_0),\mathcal{D}_n,\bss{\theta})p(\bss{\theta}\mid \mathcal{D}_n)d\bss{\theta},
\end{eqnarray}
where $\mathcal{D}_n$ denotes the sample on which the joint model was fitted and on which the predictions will be based. The first part of the integrand can be written as 
\begin{eqnarray}
	\nonumber& &\textrm{Pr}(T_i\geq j\mid T_i> j_0, \mathcal{D}_i(j_0),\mathcal{D}_n,\bss{\theta})\\
	\nonumber&= &\int \textrm{Pr}(T_i\geq j\mid T_i> j_0, \mathcal{D}_i(j_0),\bs{b}_i,\bss{\theta})p(\bs{b}_i\mid T_i> j_0, \mathcal{D}_i(j_0),\bss{\theta})d\bs{b}_i\\
	\nonumber&=& \int \textrm{Pr}(T_i\geq j\mid T_i> j_0, \bs{b}_i,\bss{\theta})p(\bs{b}_i\mid T_i> j_0, \mathcal{D}_i(j_0),\bss{\theta})d\bs{b}_i\\
	&=& \int \frac{S_i(j\mid \bs{b}_i,\bss{\theta})}{S_i(j_0\mid \bs{b}_i,\bss{\theta})}p(\bs{b}_i\mid T_i> j_0, \mathcal{D}_i(j_0),\bss{\theta})d\bs{b}_i.
\end{eqnarray}
The second part is the posterior distribution of the parameters given the observed data. By using arguments of standard asymptotic Bayesian theory and assuming that the sample size $n$ is sufficiently large, we approximate the distribution of $\{\bss{\theta}\mid \mathcal{D}_n\}$ by $N(\hat{\bss{\theta}},\hat{\Sigma})$. 

Given $\mathcal{D}_i(j_0)$ and $\bss{\theta}$, the posterior distribution of $\bs{b}_i$ is 
\begin{eqnarray}
	\nonumber	f_{\bs{b}_i}(\bs{b}_i\mid T_i> j_0,\mathcal{D}_i(j_0),\bss{\theta}) &\propto& p(T_i>j_0,\bs{b}_i,\mathcal{D}_i(j_0)\mid \bss{\theta})\\
	&=& S_i(j_0\mid \bs{b}_i, \bss{\theta})\prod_{t=1}^{j_0}f_{Y_{it}}(Y_{it}\mid \bs{b}_i,\bss{\theta})f_{\bs{b}_i}(\bs{b}_i)d\bs{b}_i.
\end{eqnarray}
This posterior distribution of the random effects is of nonstandard form.

 \abhi{Analogous to arguments presented in Rizopoulos (2011), Mclain et al. (2012)  \citep{Rizopoulos:2011, Mclain:Lum:Sundaram:2012}, we assume posterior conditional $\bs{b}_i \sim$} a multivariate $t$ distribution centered at the empirical Bayes estimates, $\hat{\bs{b}}_i=\arg\max_{\bs{b}}\{\log p(T_i>j_0,\abhi{\bs{b}},\mathcal{D}_i(j_0) \vert\hat{\bss{\theta}})\}$, and scale
matrix $$\hat{\textrm{var}}(\hat{\bs{b}}_i)=\{-\partial^2\log p(T_i>j_0,\abhi{\bs{b}},\mathcal{D}_i(j_0)\vert\hat{\bss{\theta}})/\partial\bs{b}'\partial\bs{b} \vert_{\bs{b}=\hat{\bs{b}}_i} \}\abhi{^{-1}},$$
with four degrees of freedom. 

A Monte Carlo sample of $\pi_i(j\mid j_0)$ can be obtained using the following simulation scheme:
\begin{itemize}
	\item Step 1. Draw $\bss{\theta}^{(l)}\sim N(\hat{\bss{\theta}},\hat{\Sigma})$.
	\item Step 2. Draw $\bs{b}_i^{(l)}\sim t_4\{\hat{\bs{b}}_i,\hat{\textrm{var}}(\hat{\bs{b}}_i)\}$.
	\item Step 3. Compute $\pi_i^{(l)}(j\mid j_0)=S_i\{j\mid \bs{b}_i^{(l)},\bss{\theta}^{(l)}\}/S_i\{j_0\mid\bs{b}_i^{(l)}, \bss{\theta}^{(l)}\}$.
\end{itemize}
Repeat Steps 1--3 for $l=1,\dots,L$ times, where $L$ denotes the Monte Carlo sample size. In our prediction analysis, we used $L=500$ samples to estimate the mean and $95\%$ quantile based confidence intervals. 

\section{Simulation Studies}
\label{s:simulation}
In this section, we conducted simulation studies to investigate the performance of the proposed estimates using likelihood-based approach \abhi{under various settings}. For simplicity, we assumed covariates $U_{ij}=U_i$ and generated $U_i$ from  \abhi{$\textrm{N}(2,1)$}. Let $Z_{ij}=Z_i=(1,U_i)$. $\bs{b}_i=(b_{Y,i},b_{T,i})$ were generated from  \abhi{\textbf{MVN}$(\bs{0}, D)$}, 
$$ D=\left( \begin{array}{cc}
	\sigma_1^2 & \zeta\sigma_1\sigma_2 \\
	\zeta\sigma_1\sigma_2 & \sigma_2^2
\end{array}
\right).$$
The random error $\epsilon_{ij}$ were generated from $\textrm{N}(0,0.9^2)$ and  $Y_{ij}$ were generated using (\ref{hormone_model}). \abhi{ Similarly, TTP for $i$-th subject is simulated using (\ref{TTP_model}). True values of $\bss{\beta}, \bss{\gamma}, \psi,\sigma_1,\sigma_2$, and $\zeta$ are listed in Table~\ref{tab1}}. Subjects who had not experienced an event \abhi{till} $j=6$ were censored.

\abhi{We have considered 9 different simulation settings with varying sample sizes ($n$= 300, 400, and 500) as well as, with varying intercourse success probabilities ($p(A_{ij})$= 0.95,0.90 and 0.85). These choices are made in such a way that a particular case ($n=400$, $p(A_{ij})=0.95$) closely mimics the real data in terms of distribution of TTP while the other cases either overestimate or under-estimate the number of cycles and sample size in comparison to real data. For each simulation scenario, $m=$1000 replicates are generated. We split each replicate data into a training set (randomly selected 2/3rd) and  a test set, as is originally only done for real data.  Table \ref{tab0} provides a summary of TTP distribution (both overall as well as in the training set). The shaded row indicates the simulation scenario that mimics the real data in terms of sample size ($337$ in real data),  \# of cycles ($1023$ in real data), intercourse probability ($\approx 94.7\%$ in real data) and \# of cycles in training data ($686$ in real data). In Table \ref{tab0} we have provided avg. TTP,  proportion of right censoring ($\hat{p}(\delta=0)$), \# of cycles in overall as well as in training sets, each averaged (mean(Mn)/median(Med)) over 1000 replicates (provided with interquartile range (IQR)).

\begin{table}[ht]  
	\centering
	\begin{minipage}{\textwidth}
	\caption{Summary of TTP distribution across simulation settings: avg. TTP, \# of cycles and prop. of right censoring ($\hat{p}(\delta=0)$) averaged over 1000 replicates.The highlighted row mimics the oxford data.}    \label{tab0}  
	\small
	\begin{tabular}{@{\extracolsep{\fill}}lccccc@{\extracolsep{\fill}}}
		\toprule
		\multicolumn{2}{c}{Settings} & \multicolumn{3}{c}{ TTP (full data)} & \multicolumn{1}{c}{ TTP (training set)} \\\hline
		&  & Avg. TTP & \# of cycles & $\hat{p}(\delta=0)$ & \# of cycles \\
		$n$     &  $p(A_{ij})$    &  Mn (IQR)  & Med (IQR)   & Mn (IQR)  & Med (IQR) \\\hline
		300 & 0.95 & 2.83 (0.16)  & 0848 (47)  & 0.24 (0.03)  & 567 (38) \\
		300 & 0.90 & 2.88 (0.16)  & 0862 (47)  & 0.24 (0.03)  & 576 (38) \\
		300 & 0.85 & 2.93 (0.16)  & 0877 (47)  & 0.24 (0.03)  & 586 (39) \\
		\shade{400} & \shade{0.95} & \shade{2.83 (0.14)}  & \shade{1133 (56)}  & \shade{0.24 (0.03)}  & \shade{755 (45)} \\
		400 & 0.90 & 2.87 (0.14)  & 1148 (57)  & 0.24 (0.03)  & 767 (43) \\
		400 & 0.85 & 2.92 (0.14)  & 1170 (56)  & 0.24 (0.03)  & 781 (45) \\
		500 & 0.95 & 2.83 (0.12)  & 1416 (62)  & 0.24 (0.03)  & 945 (51) \\
		500 & 0.90 & 2.88 (0.12)  & 1438 (60)  & 0.24 (0.03)  & 960 (49) \\
		500 & 0.85 & 2.92 (0.12)  & 1462 (62)  & 0.24 (0.03)  & 975 (49) \\
		\botrule
	\end{tabular}
\end{minipage}
\end{table}

\bmhead{Estimate accuracy with the training sets:} 
\noindent For each replicate data, training set is used to fit the model,  while test set is used to estimate predictive accuracy. Based on training set, we have reported estimation bias (Bias), standard deviation (SD) and coverage probability (CP) based on 95\% CI of the parameter estimate with varying samples sizes ($n$= 300, 400, and 500) for a given intercourse success probability, p($A_{ij}$)= 0.95,  in Table \ref{tab1}. Same set of  metrics are reported for p($A_{ij}$)= 0.90 and 0.85 in Appendix~\ref{AppTables}, Table~\ref{tab2},  and Table~\ref{tab3} respectively. $n_{tr}$ represents the training set size, which is 2/3rd of the sampled data.} Estimation used Gaussian quadrature with 50 nodes and the simulation was conducted in software \texttt{R/4.2}.
\begin{table}[ht]  
	\centering
	\begin{minipage}{\textwidth}
	\caption{Simulation results corresponding to varying sample sizes with $p(A_{ij})=0.95$}    \label{tab1}  
	\small
	\begin{tabular}{@{\extracolsep{\fill}}l@{\extracolsep{\fill}}c@{\extracolsep{\fill}}|ccl|@{\extracolsep{\fill}}ccc|@{\extracolsep{\fill}}ccc@{\extracolsep{\fill}}}
		\toprule
		\multicolumn{2}{@{\extracolsep{\fill}}c@{\extracolsep{\fill}}}{Param.} & \multicolumn{3}{c@{\extracolsep{\fill}}}{n=300, $n_{tr}$=200} & \multicolumn{3}{c}{n=400, $n_{tr}$=267} & \multicolumn{3}{c}{n=500, $n_{tr}$=334} 
		\\\hline
		Sym.  &True & Bias  & SD & CP & Bias  & SD & CP & Bias  & SD & CP \\\hline
		$\beta_1$ & 4 & 0.002 & 0.13 & 0.95 & -0.004 & 0.111 & 0.941 & -0.005 & 0.103 & 0.952\\
		$\beta_2$ & -0.5 & -0.004 & 0.049 & 0.957 & 0.001 & 0.043 & 0.947 & 0.001 & 0.04 & 0.951\\
		$\gamma_1$ & -27 & -0.182 & 1.538 & 0.976 & -0.171 & 1.738 & 0.97 & -0.041 & 1.423 & 0.963\\
		$\rho_1$ & -6.5 & 0.349 & 2.239 & 0.961 & 0.503 & 2.257 & 0.958 & 0.521 & 1.852 & 0.964\\
		$\rho_2$ & 10 & 0.49 & 2.052 & 0.961 & 0.437 & 1.965 & 0.967 & 0.482 & 1.627 & 0.971\\
		$\rho_3$ & 17 & 0.462 & 1.855 & 0.929 & 0.457 & 1.904 & 0.937 & 0.474 & 1.64 & 0.919\\
		$\rho_4$ & 21 & 0.525 & 1.859 & 0.96 & 0.503 & 1.959 & 0.964 & 0.472 & 1.577 & 0.929\\
		$\rho_5$ & 24 & 0.525 & 1.837 & 0.968 & 0.621 & 1.986 & 0.969 & 0.516 & 1.643 & 0.949\\
		$\rho_6$ & 25 & 0.293 & 2.131 & 0.971 & 0.385 & 2.086 & 0.965 & 0.411 & 1.721 & 0.962\\
		$\psi_{\mu}$ & 18 & 0.044 & 1.48 & 0.972 & 0.01 & 1.645 & 0.973 & -0.091 & 1.353 & 0.974\\
		$\sigma_1$ & 0.3 & 0.005 & 0.069 & 0.929 & 0.007 & 0.065 & 0.916 & 0.006 & 0.056 & 0.926\\
		$\sigma_2$ & 3 & 0.587 & 2.109 & 0.956 & 0.53 & 2.003 & 0.942 & 0.464 & 1.813 & 0.949\\
		$\sigma$ & 0.9 & -0.004 & 0.03 & 0.947 & -0.003 & 0.028 & 0.933 & -0.002 & 0.025 & 0.944\\
		$\zeta$ & -0.2 & 0.008 & 0.469 & 0.921 & 0 & 0.429 & 0.906 & 0.004 & 0.381 & 0.929\\
		\botrule
	\end{tabular}
\end{minipage}
\end{table}
The simulation results show that the proposed estimation approach works well in all settings considered here with reasonably small bias and coverage probabilities close to the nominal level. \abhi{For $\rho_3$, $\sigma_1$,  and $\zeta$, CP falls slightly below 95\% relative to other parameters, which may be attributable to numerical approximation of marginal likelihood.}

\abhi{
\paragraph{Prediction accuracy with the test sets:}
We have  implemented the prediction methodology analogous to Rizopoulos (2011)\cite{Rizopoulos:2011} as described in Section~\ref{s:estimation}. In particular, we have estimated  subfertility probability $\pi_i{(6 \mid 1)}=p(TTP_i > 6 \mid TTP_i >1)$ for the $i$-th test sample given that $TTP_i > 1$ and used that to see how it performs as prediction rule to determine if this sample is right censored or not at sixth cycle. We have used standard ROC plot and area under the curve (auc) as used for real data by varying cutoff points, and computing empirical sensitivities and specificities (See Section~\ref{s:estimation}  for more details). To determine how good the ROC plot and auc estimates are, we have calculated quantile-based 95\% CI for auc and 95\% CI band for ROC plots. The auc estimates are summarized in below table along with the effective test sample sizes used. The effective test sample size, $n_e$, represents the subset for which $TTP_i >1$. In the below table $n_{tst}$ indicates the test sample size initially considered, which is 1/3rd of $n$. 
Table \ref{tab4} indicates the decision rule is extremely effective in predicting subfertility.  
\begin{table}[ht!]  
	\centering
	\caption{Summary: auc and effective  test sample size ($n_{e}$) based on 1000 test data replicates}    \label{tab4}  
	\small
	\begin{tabular}{|@{\extracolsep{\fill}}c@{\extracolsep{\fill}}|@{\extracolsep{\fill}}c@{\extracolsep{\fill}}c@{\extracolsep{\fill}}|@{\extracolsep{\fill}}c@{\extracolsep{\fill}}c@{\extracolsep{\fill}}|@{\extracolsep{\fill}}c@{\extracolsep{\fill}}c@{\extracolsep{\fill}}|}
		\hline
		& \multicolumn{2}{c}{n=300, $n_{tst}$=100} & \multicolumn{2}{c}{n=400, $n_{tst}$=133} & \multicolumn{2}{c|}{n=500, $n_{tst}$=166} 
		\\\hline
		$p(A_{ij})$ & auc (95\% CI) &  $n_{e}$(IQR) & auc (95\% CI) &  $n_{e}$(IQR)  & auc (95\% CI) &  $n_{e}$(IQR) \\\hline
		0.95 & 0.983 (0.95, 1) & 59 (7) & 0.983 (0.96, 1)  & 78 (8) & 0.983 (0.96, 1)  & 98 (9) \\
		0.90 & 0.984 (0.95, 1) & 61 (6) & 0.984 (0.96, 1)  & 80 (7)  & 0.984 (0.96, 1)  & 101 (8) \\
		0.85 & 0.984 (0.95, 1) & 63 (6) & 0.984 (0.96, 1)  & 84 (7) &  0.985 (0.96, 1)  & 105 (8) \\
		\hline
	\end{tabular}
\end{table}

 To get a sense of predictive performance we have overlayed 1000 ROC plots each from a replicate. While constructing  quantile-based 95\% CI may be straightforward, constructing a 95\% confidence band for ROC function may not be so.  Let sensitivity be $f(c)$ for a given cutoff $c$. Similarly, let $g(c)$ be the 1-specificity function. ROC function can now be defined as $f o g^{-1}$ assuming $f$ and $g$ are smooth one-one functions. Since empirical sensitivity and specificity functions are step functions, thus neither smooth nor one-one we approximate $g^{-1}$ by smooth quantile function just to guarantee all ROC functions from various replicates correspond to the same cutoff points while being overlayed. Note that the sensitivity function (therefore the ROC function) is not affected by the approximation on $g^{-1}$ if sufficiently fine quantiles of specificity functions are used while plotting. The mean ROC curves and 95\% CI bands are then computed point-wise. It is noteworthy that the area under the mean ROC curve and 95\% upper and lower confidence curves may not be exactly the same as with the mean and 95\%  CI's of  auc's as computed in Table \ref{tab4} but they are expected to be close to each other as is the case.  The mean ROC curve with 95\% band is given below in Figure \ref{ROC_sim_plots} for each of the 9 cases.  From these plots, it is evident that the prediction methodology maintains high accuracy in all simulation settings.
}
\begin{figure}[!hbt]
	\begin{subfigure}{0.32\textwidth} 
		\includegraphics[scale=0.5,width=\linewidth]{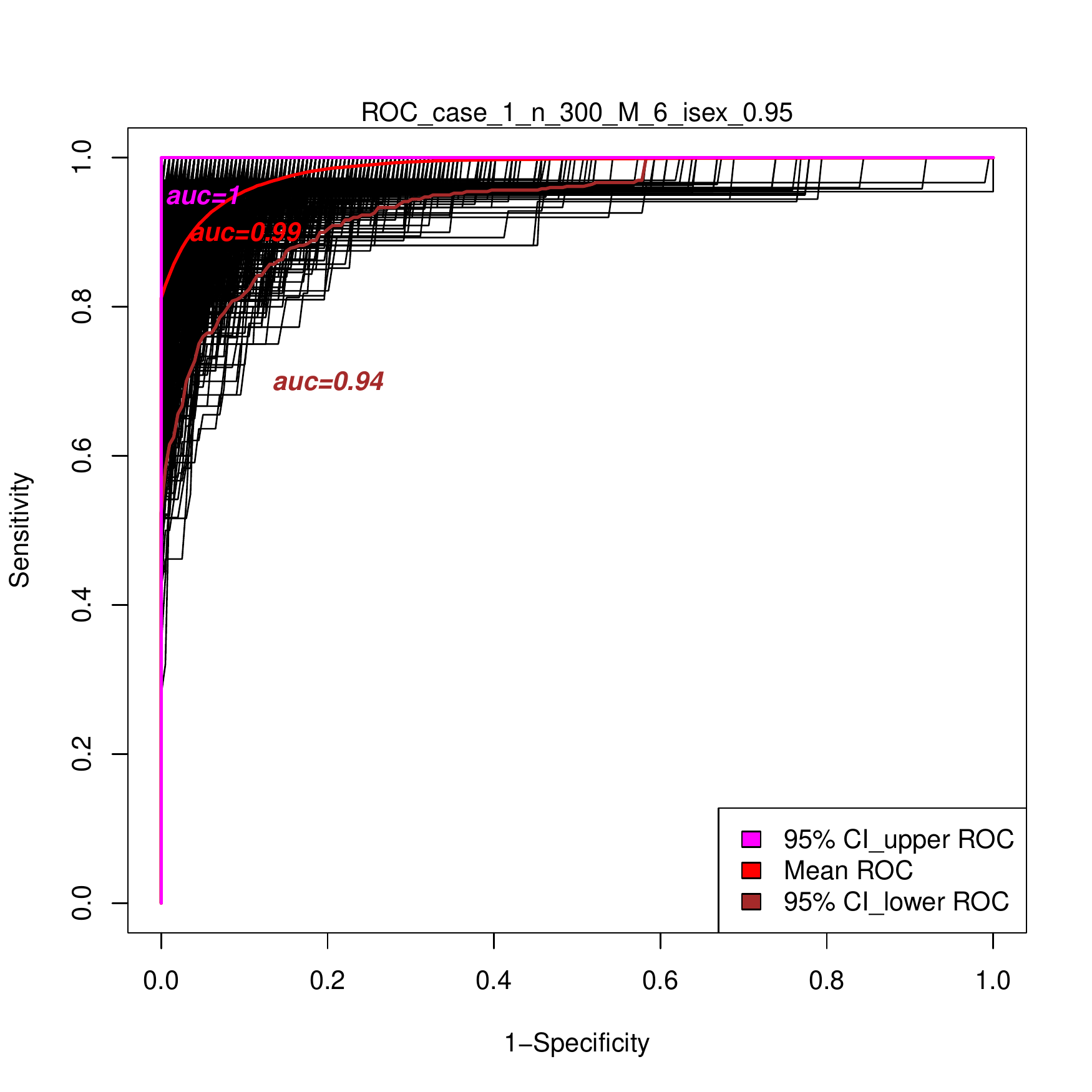}    
	\caption{ \small $n_{tst}$=100, p(A$_{ij}$)=0.95}  \label{case1} 	\end{subfigure}
	\begin{subfigure}{0.32\linewidth}
		\includegraphics[width=\linewidth]{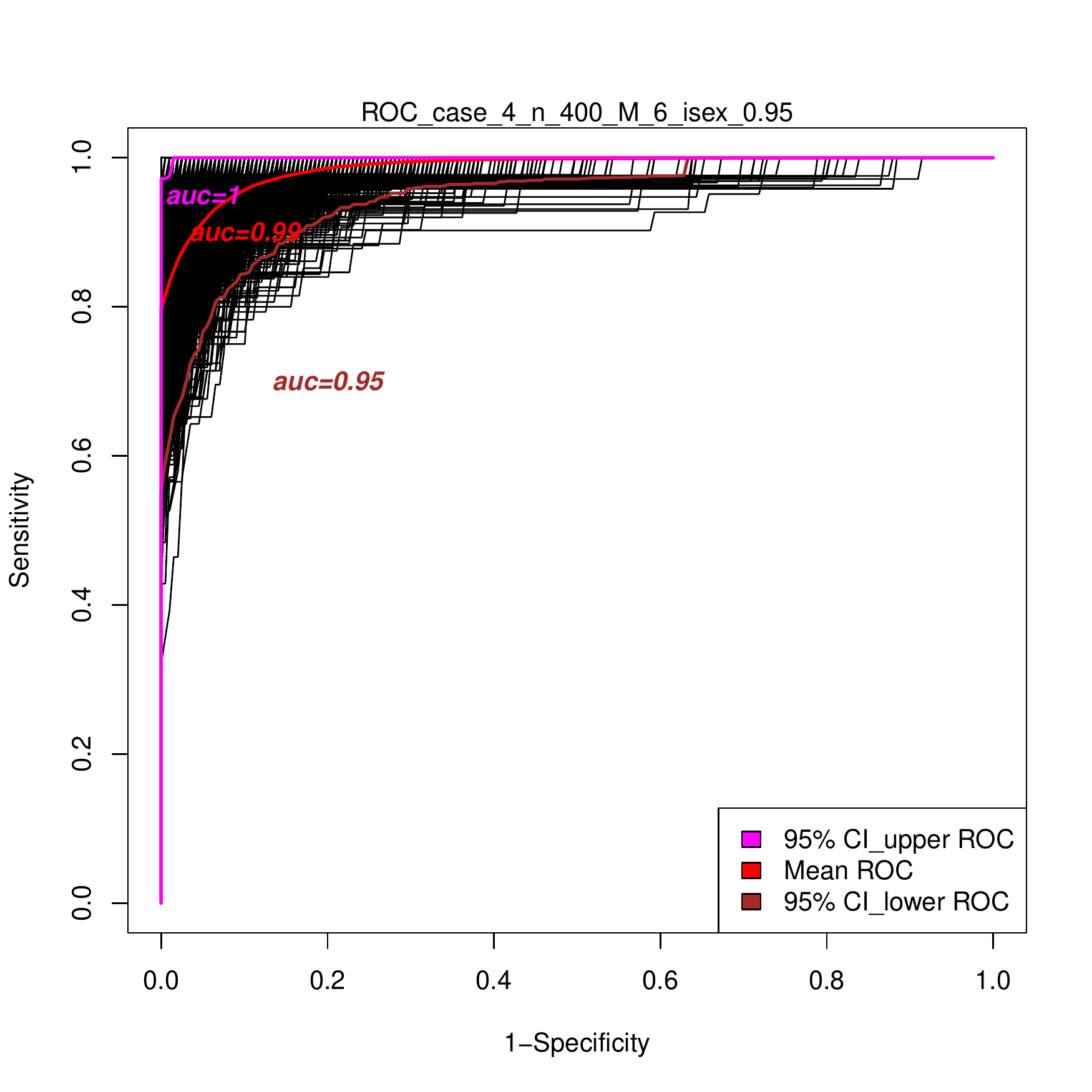}   		\caption{$n_{tst}$=133, p(A$_{ij}$)=0.95}\label{case4}
	\end{subfigure} 
	\begin{subfigure}{0.32\linewidth}
		\includegraphics[width=\linewidth]{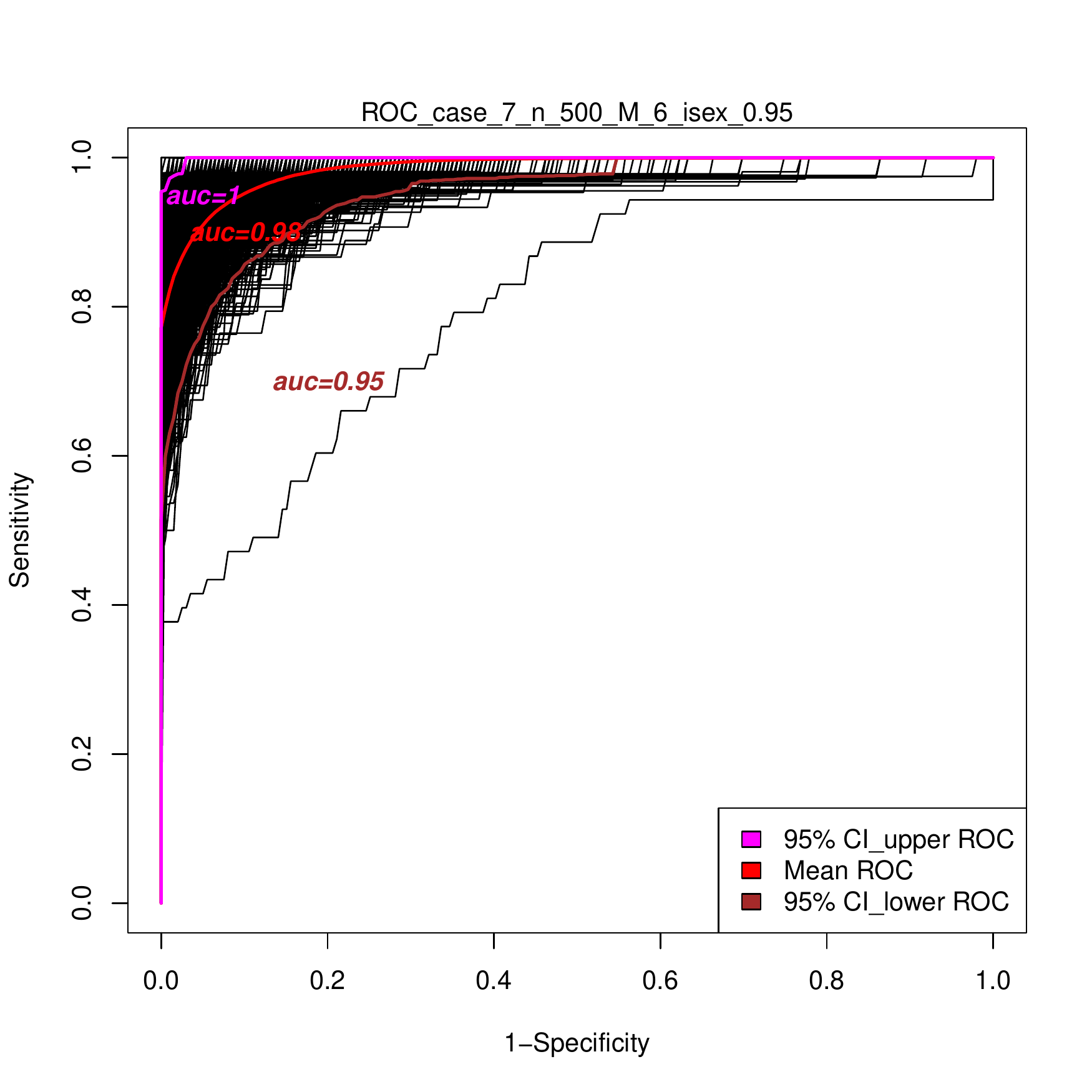}    \caption{$n_{tst}$=166, p(A$_{ij}$)=0.95}\label{case7}
	\end{subfigure} \\
	\begin{subfigure}{0.32\linewidth} 
		\includegraphics[width=\linewidth]{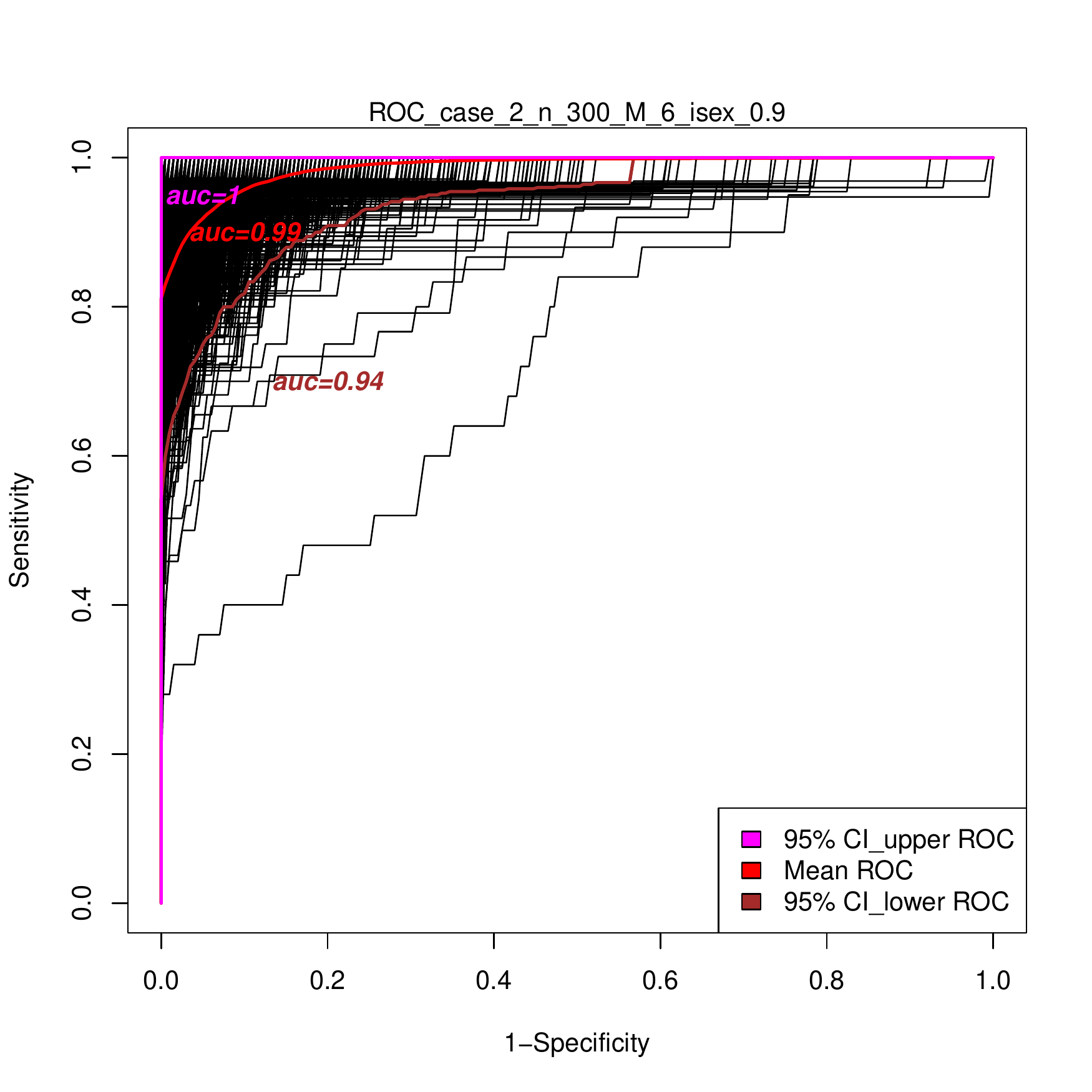}    \caption{ \small $n_{tst}$=100, p(A$_{ij}$)=0.90}  \label{case2} 
	\end{subfigure}
	\begin{subfigure}{0.32\linewidth}
		\includegraphics[width=\linewidth]{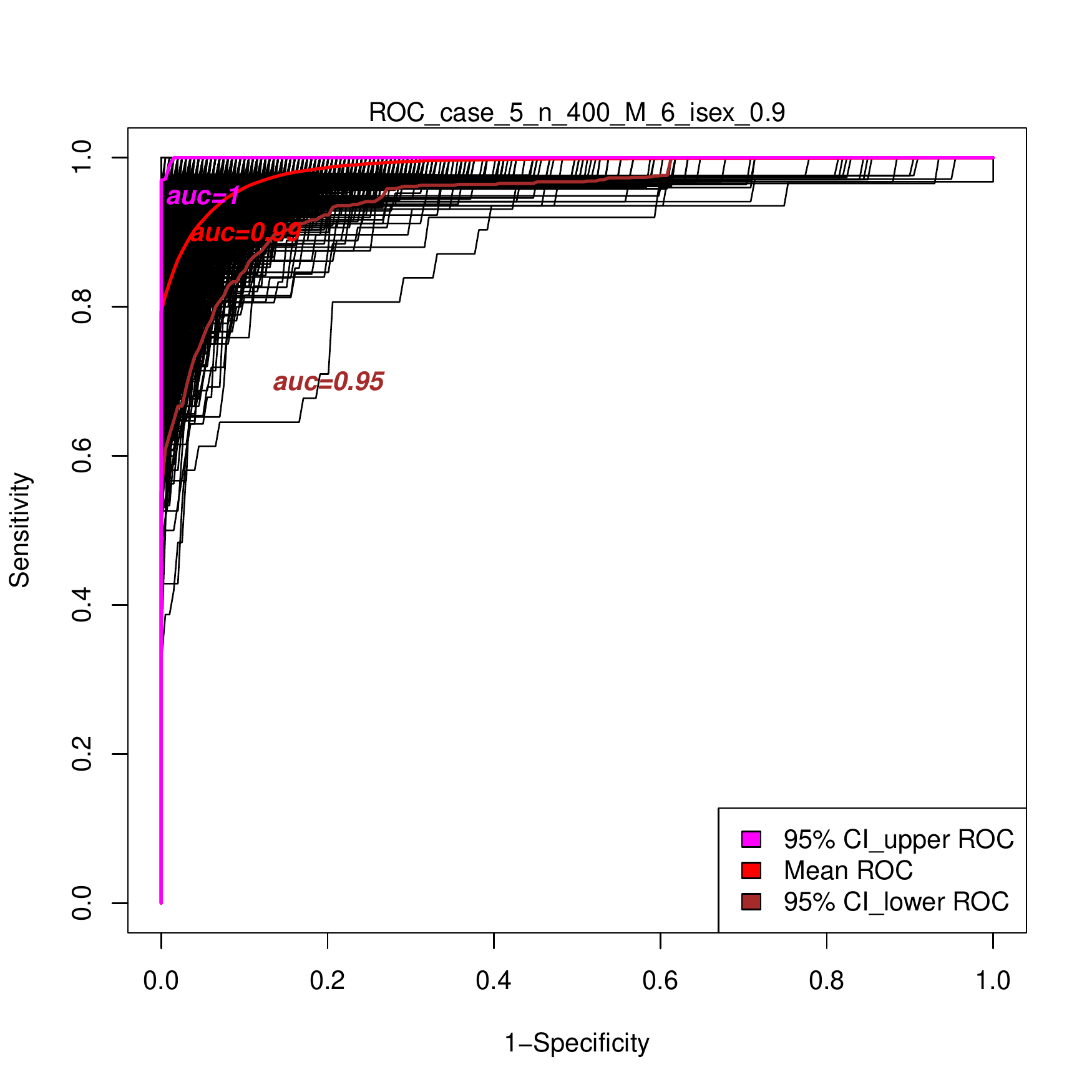}   
		\caption{$n_{tst}$=133, p(A$_{ij}$)=0.90}\label{case5}
	\end{subfigure} 
	\begin{subfigure}{0.32\linewidth}
		\includegraphics[width=\linewidth]{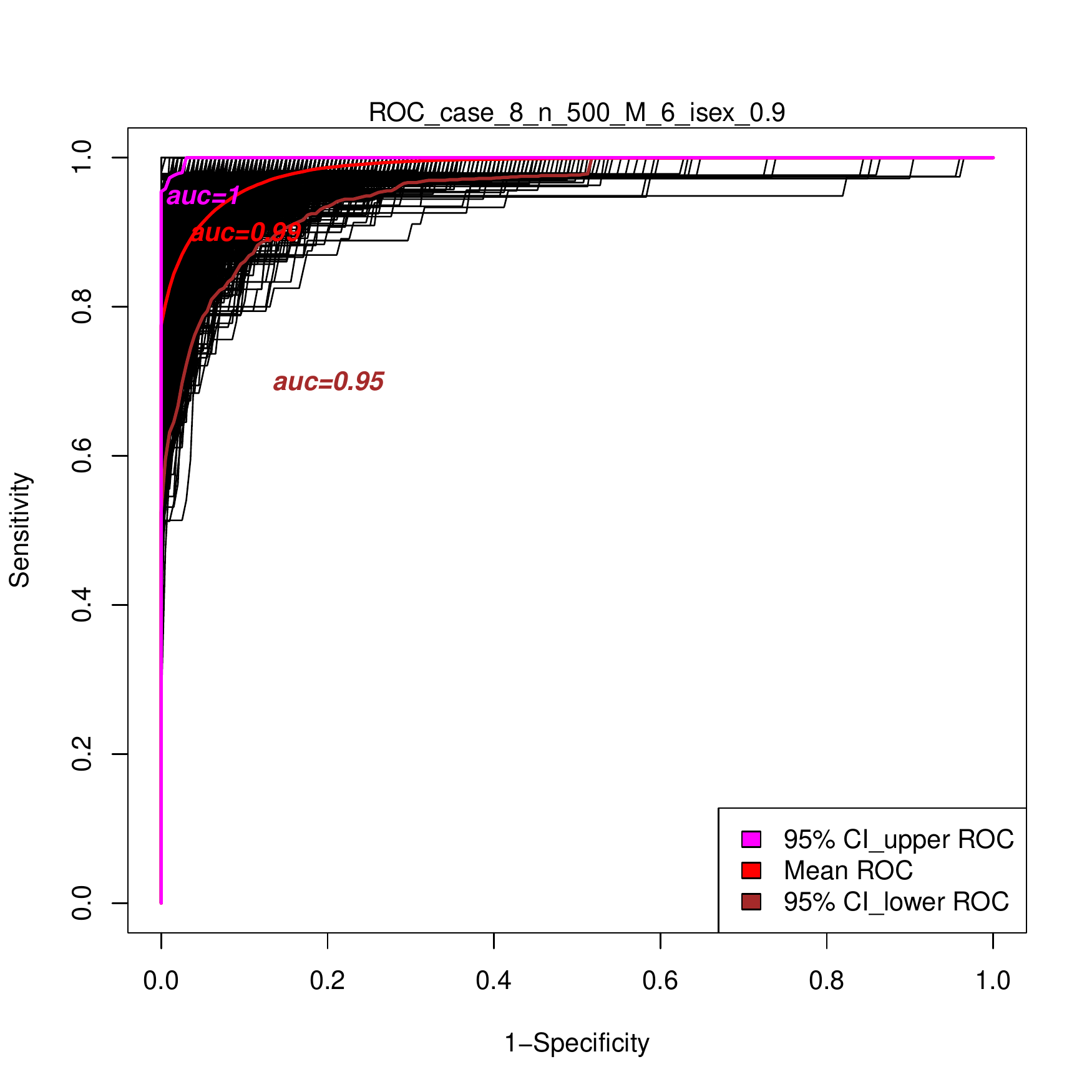}    \caption{$n_{tst}$=166, p(A$_{ij}$)=0.90}\label{case8}
	\end{subfigure}\\
	\begin{subfigure}{0.32\linewidth} 
		\includegraphics[width=\linewidth]{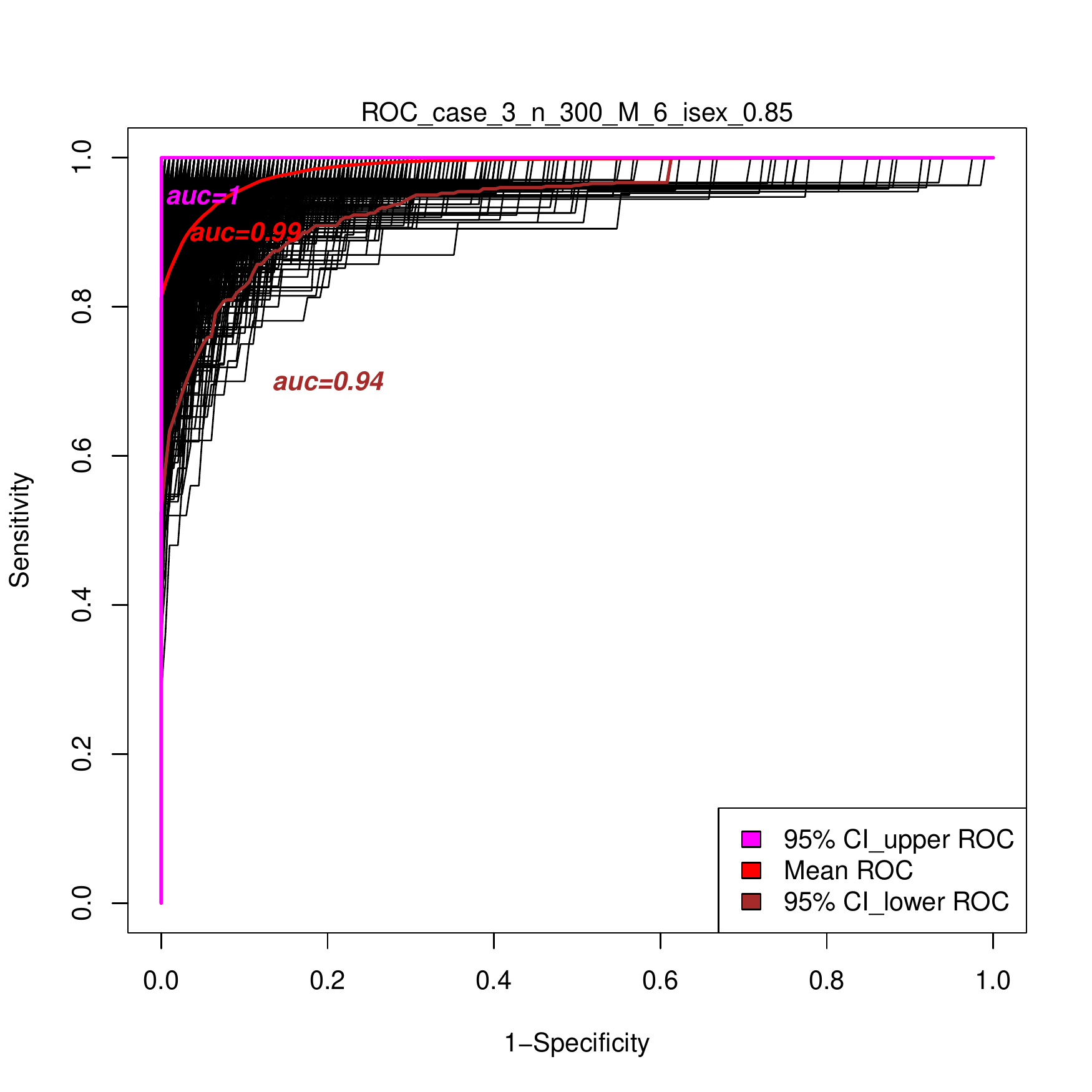}    \caption{ \small $n_{tst}$=100, p(A$_{ij}$)=0.85}  \label{case3} 
	\end{subfigure}
	\begin{subfigure}{0.32\linewidth}
		\includegraphics[width=\linewidth]{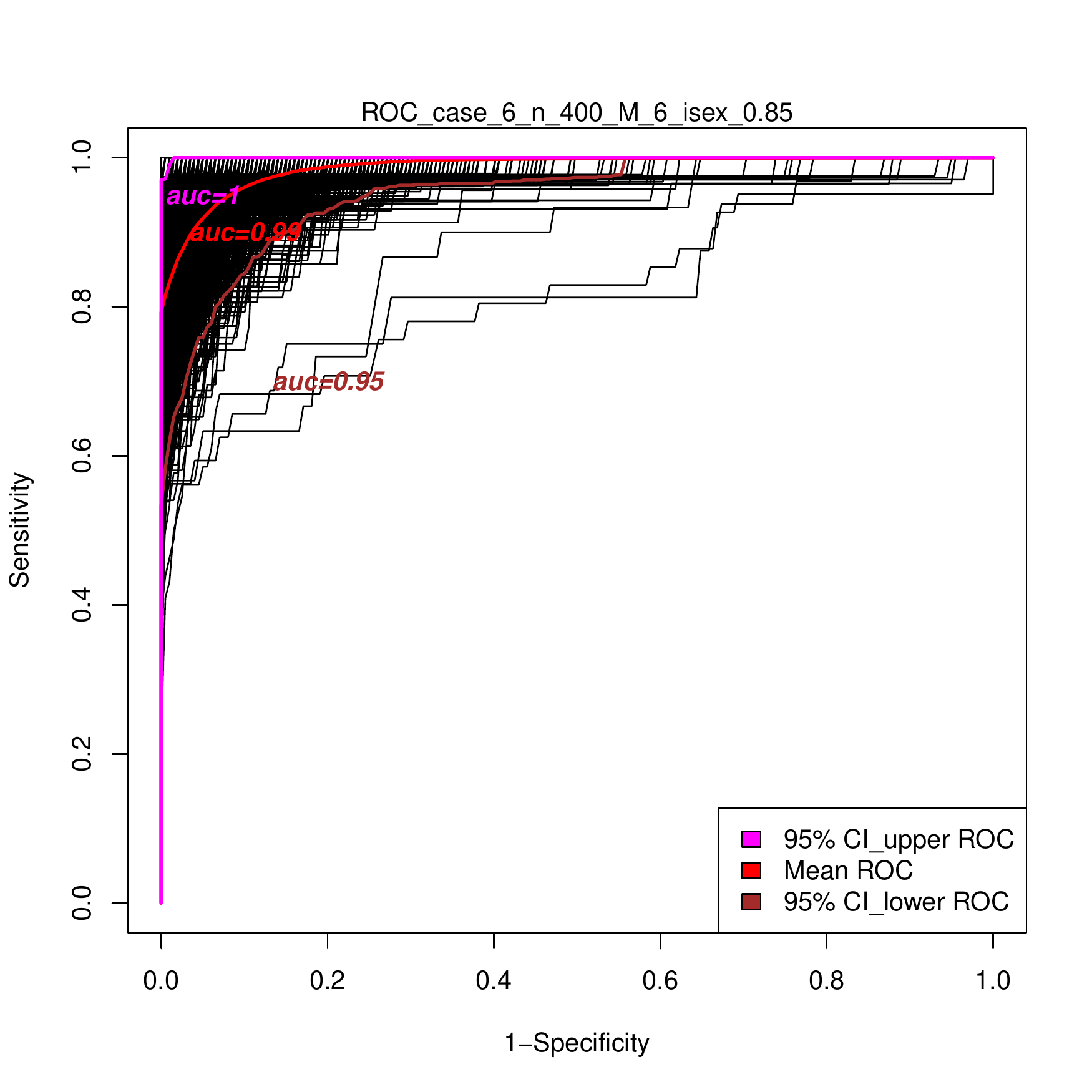}   \caption{$n_{tst}$=133, p(A$_{ij}$)=0.85}\label{case6}
	\end{subfigure} 
	\begin{subfigure}{0.32\linewidth}
		\includegraphics[width=\linewidth]{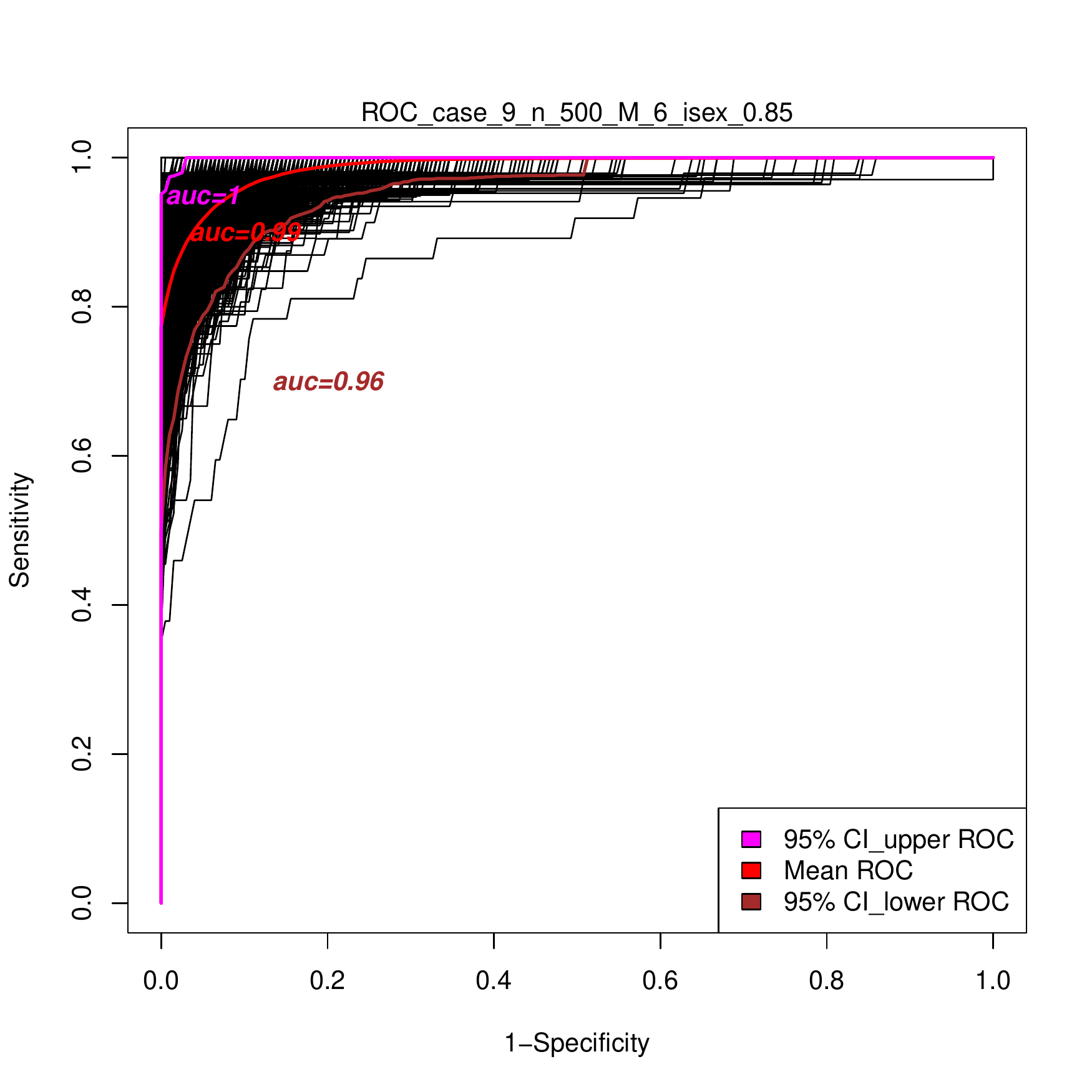}    \caption{$n_{tst}$=166, p(A$_{ij}$)=0.85}\label{case9}
	\end{subfigure}
	\caption{ROC plots with mean function and 95\% CI  band}\label{ROC_sim_plots}
\end{figure}
\abhi{
\begin{remark}{\textbf{Simulating longitudinal data at cycles:}}
We would like to bring reader's attention to the fact that unlike the real data applications, the geometric features are directly generated from the linear model.  This way of simulating geometric feature  is equivalent to summarizing simulated longitudinal data under certain simulation schemes, and the same method can work for any geometric feature. This point is elaborated in Appendix \ref{Long}.  
\end{remark}
}
\section{Analysis of the Oxford Conception Study}
\label{s:realdata}

We illustrate our proposed method by analyzing the Oxford Conception Study, Stress and Time-to-Pregnancy component \citep{Pyper:etal:2006}, which is a prospective cohort study with preconception enrollment of women aged 18-40 years who were attempting to become pregnant.  The women in the Oxford Conception Study (hereafter OCS) provided daily level information on reproductive hormones, in particular, luteinizing hormone during mid-cycle, intercourse acts, and host of other lifestyle variables, along with couple level baseline covariates. The lutenizing hormone was observed through the ovulation kits used by the woman to track her daily fertility level to identify ovulation, the values provided by the monitor were monotonically transformed values of the lutenizing hormones (manufacturer only provides these transformed values for research purpose). These women were prospectively followed for the number of menstrual cycles it took them to become pregnant (i.e., human-chorionic gonadotropin confirmed pregnancy on the day of expected menses), i.e., TTP, or a maximum of $6$ menstrual cycles.  Other examples of prospective pregnancy cohort studies' include the Longitudinal Investigation of Fertility and Environment \citep{Louis:etal:2011}, Fertili \citep{Colombo:Masarotto:2000}, and Billings \citep{Colombo:etal:2006}. 

\begin{figure}
	\centering
	\includegraphics[width=0.45\textwidth]{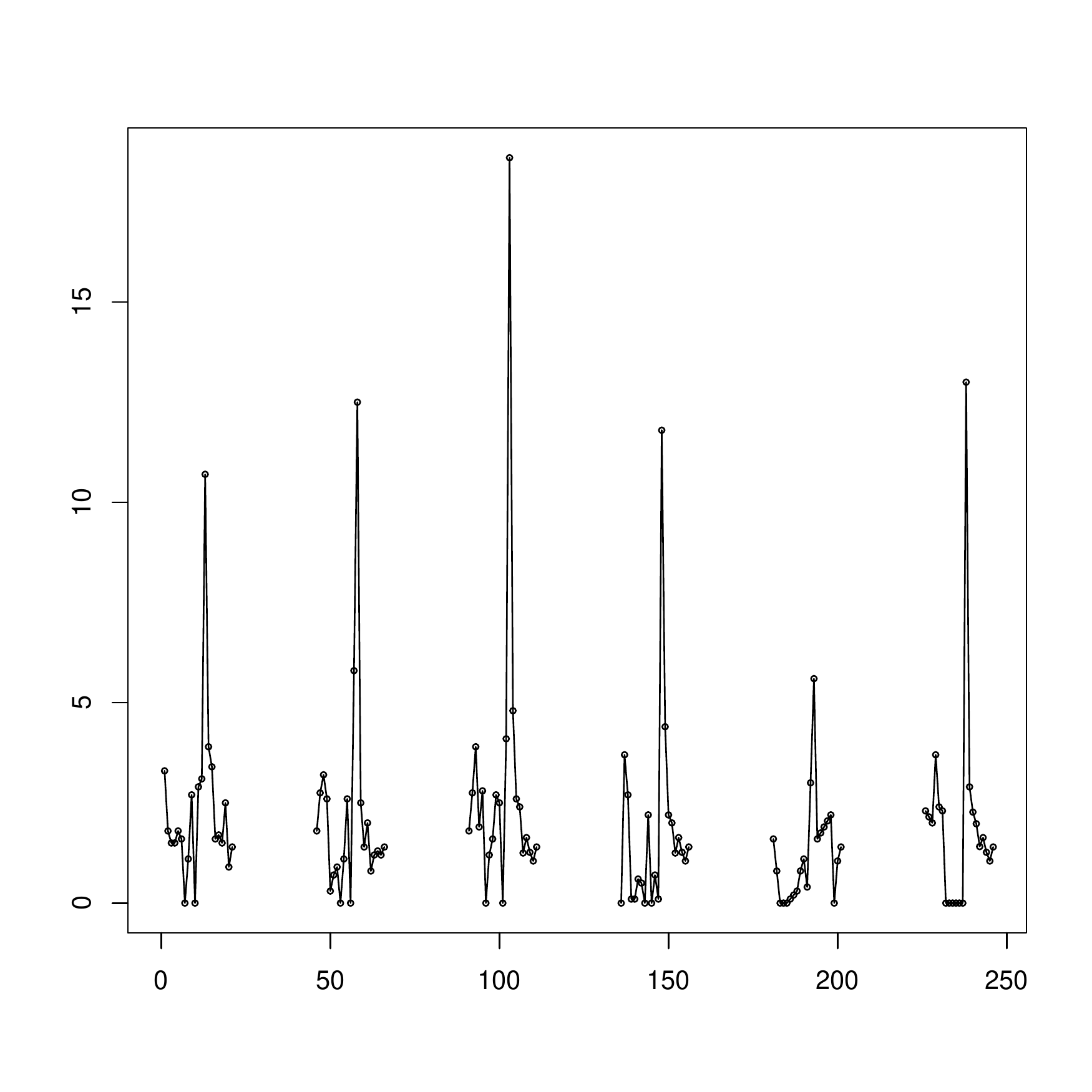}
	\includegraphics[width=0.45\textwidth]{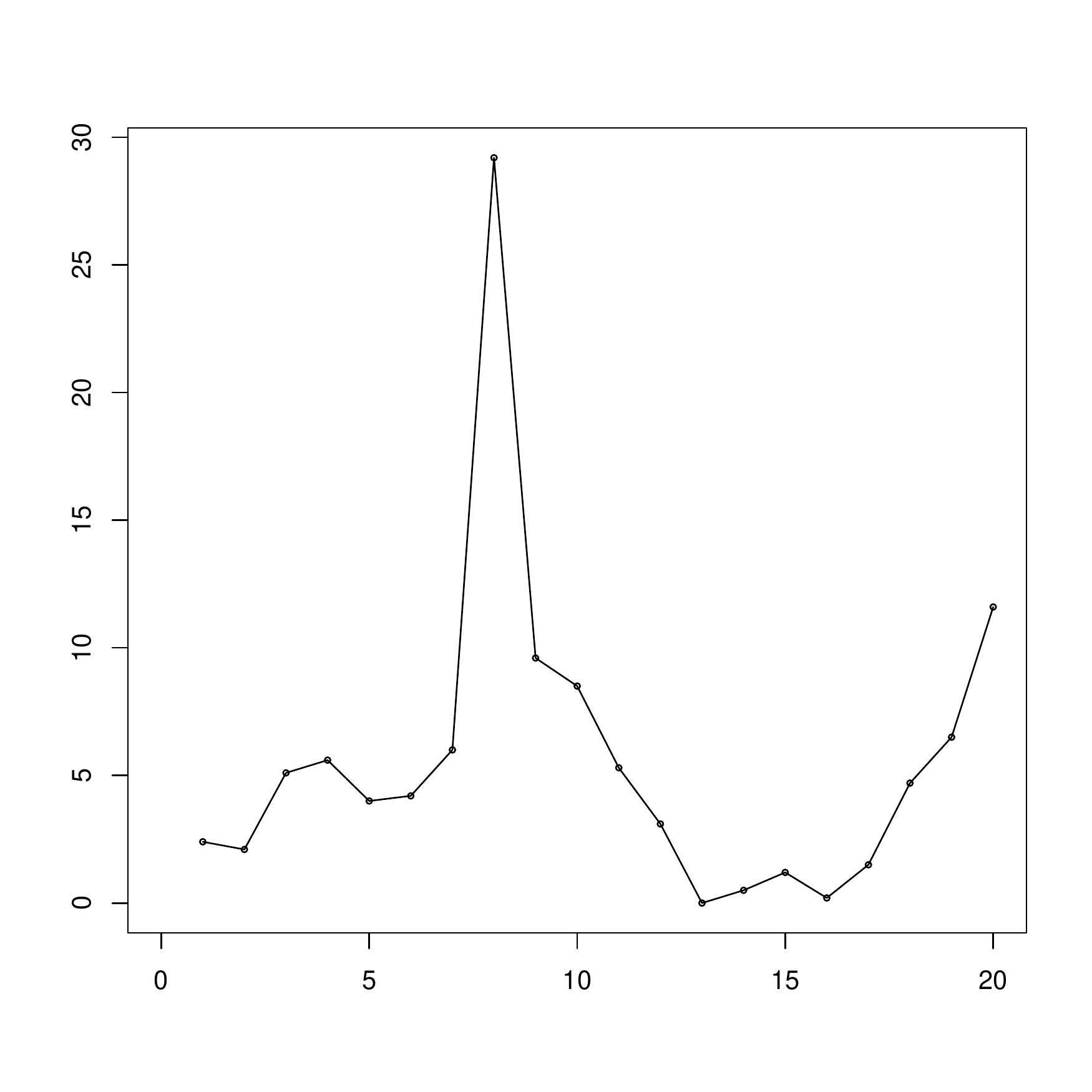}\\
	\includegraphics[width=0.45\textwidth]{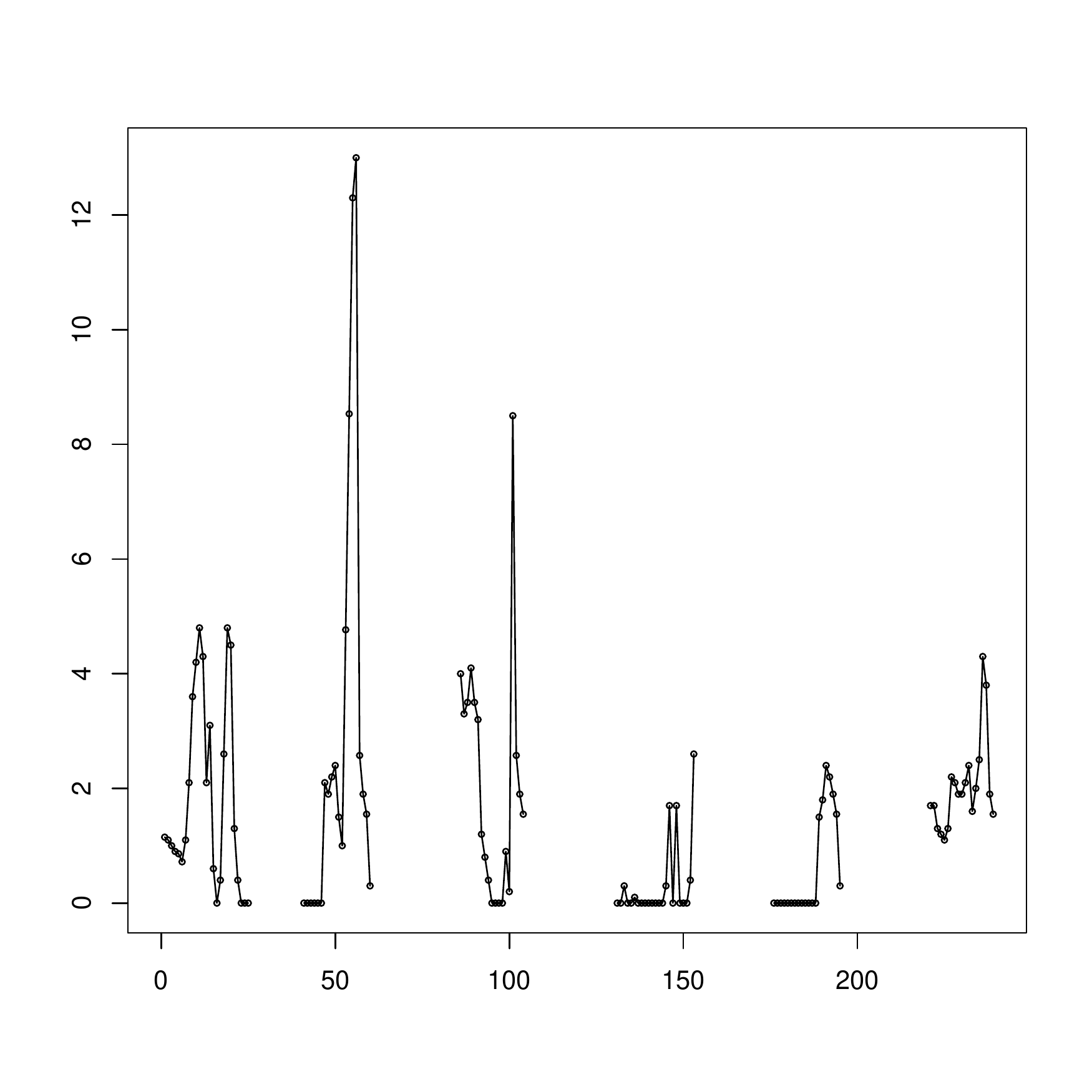}
	\includegraphics[width=0.45\textwidth]{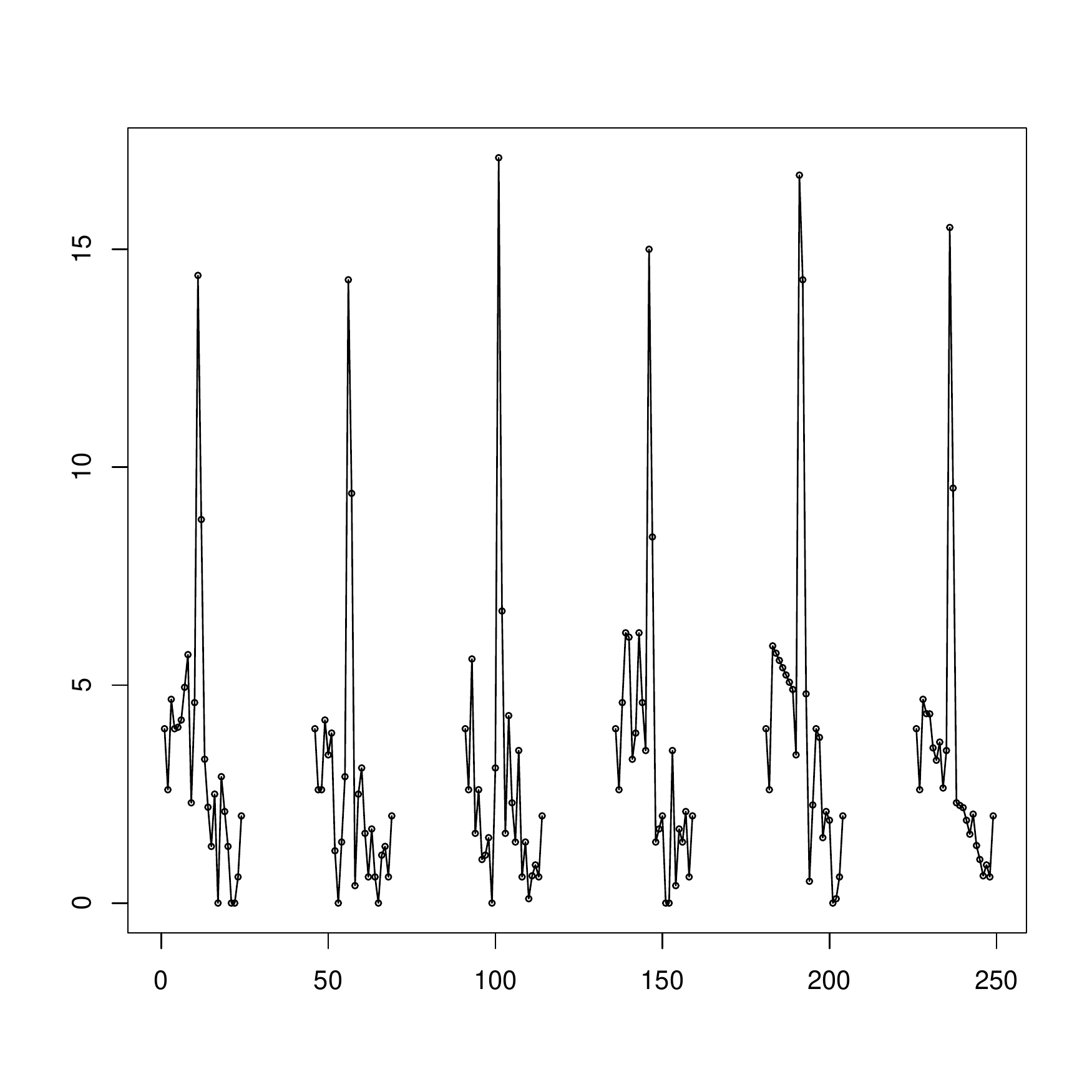}
	\caption{LH measurements profile curves by days for 4 randomly selected subjects in the data set.}
	\label{fig:LHprofile}	
\end{figure}

The data consisted of  337 women with LH measurements, resulting in a total of 1023 menstrual cycles. We randomly selected 225 (about two-thirds) of the women  with 686 cycles in the training set and the rest of the women were taken to be the prediction set. The hormonal measurements were taken daily within a fixed window in each menstrual cycle. For each cycle of every woman, we used B-spline functions to smooth the observed hormone data and then calculate the geometric features of interest. As mentioned in previous sections, we considered three geometric features of the hormone profiles: curvature at the LH peak which measures the sharpness of the LH profile at peak, LH peak value and the average LH profile curvature within fertile window. Figure~\ref{fig:LHprofile} gives an example of the LH measurements for 4 randomly selected women in the data set.
Note the number of varying cycles worth of information based on TTP for each woman in the figure. Motivated by the underlying biological hypothesis discussed in detail in the introduction, we fit three different joint models corresponding to each of the geometric feature. In each joint model, we model TTP with each one of the cycle-level geometric features using the training dataset and evaluate their prediction abilities, respectively. In our analysis, we focus on the following covariates: female age (in model for hormonal data), couple average age and difference between female and male age (in model for TTP), and female's body mass index (BMI) (categorized as underweight or normal weight if BMI$<25$; overweight if $25\leq$BMI$<30$; obese if $30\leq$BMI)), female's smoking category (smoke.n if not smoking; smoke.m if smoking but average cigarette smoked per day $\leq 10$; smoke.h if average cigarette smoked per day $> 10$), stress level measured by the salivary biomarker alpha amylase and parity (nulliparous or multiparous; nulliparous women are those who haven't had live birth before) in both models. The three age related covariates as well as alpha amylase level were scaled by suitable constants in the analysis and were denoted as Age*, avg.age*, dif.age* and Alp*, respectively. The geometric features of hormonal profiles were also scaled by adequate constants in the analysis. Subjects with BMI smaller than 25 who are non-smokers and nulliparous were considered as the reference group.

Table~\ref{t:kpeak_Joint} presents the estimates and $95\%$ confidence intervals for the covariate effects on the curvature at LH peak, covariate effects on TTP, cycle-specific baseline effect on TTP, effect of curvature at LH peak on TTP as well as the variance covariance parameters for the random effects. Notice that all covariates seem non-significant for the curvature at LH peak. A couple's average age and female-male age difference seem to have significant negative effect on TTP. Women who smoke but smoke no more than 10 cigarettes per day have significantly longer TTP compared to non-smoking women while the difference between the heavy smoking women ($> 10$ cigarettes per day) and non-smoking women was not significant. Multiparous women seem to have significantly shorter TTP than nulliparous women. {\it We also found that the curvature at LH peak has significant positive effect on TTP.} This implies that the sharper a woman's LH peak is, the shorter her TTP tends to be. 

\begin{table}
	\caption{Joint model estimation results on curvature at LH peak.}
	\label{t:kpeak_Joint}
	\centering
	\fbox{
		\begin{tabular}{rrrr}
			Parameter& Estimate ($95\%$CI) & Parameter& Estimate ($95\%$CI) \\\midrule 
			$\beta$: Age*& $-$0.51 ($-$2.73, 1.72)&$\gamma$: avg.age*& $-$27.89 ($-$42.76, $-$13.02) \\
			$\beta$: Age*$^2$& 0.09 ($-$0.62, 0.80)&$\gamma$: dif.age*& $-$5.17 ($-$8.30, $-$2.03) \\
			$\beta$: Overweight& 0.04 ($-$0.17, 0.25) &$\gamma$: Overweight& $-$2.99 ($-$7.49, 1.51) \\
			$\beta$: Obese& $-$0.19 ($-$0.45, 0.07) &$\gamma$: Obese& 5.66 ($-$0.20, 11.51)\\
			$\beta$: Smoke.m& 0.02 ($-$0.21, 0.25) &$\gamma$: Smoke.m& $-$7.59 ($-$13.45, $-$1.72) \\
			$\beta$: Smoke.h& 0.05 ($-$0.46, 0.55)&$\gamma$: Smoke.h& $-$5.23 ($-$15.15, 4.69) \\
			$\beta$: Alp*& $-$0.15 ($-$0.47, 0.18) &$\gamma$: Alp*& $-$3.50 ($-$9.82, 2.81) \\
			$\beta$: Alp*$^2$& 0.02 ($-$0.07, 0.12)&$\gamma$: Alp*$^2$ & 1.80 ($-$0.16, 3.76) \\
			$\beta$: Parity& $-$0.08 ($-$0.27, 0.10)&$\gamma$: Parity& 9.11 (3.10, 15.13) \\
			$\rho_1$& $-$6.49 ($-$16.49, 3.50)& $\rho_2$& 10.01 (1.77, 12.25) \\
			$\rho_3$& 17.71 (14.96, 20.47)& $\rho_4$& 21.03 (16.41, 25.65) \\
			$\rho_5$& 24.36 (18.25, 30.47)& $\rho_6$& 24.96 (18.06, 31.86) \\
			$\rho_7$& $-$2.94 ($-$2.94, $-$2.94)& $\psi_\mu$& 18.63 (3.93, 33.33) \\
			$\sigma_1$& 0.35 (0.25, 0.50)& $\sigma_2$& 30.44 (18.23, 50.83) \\
			$\sigma$& 0.96 (0.90, 1.02)& $\zeta$& $-$0.20 ($-$0.31, $-$0.08) \\
	\end{tabular}}	
\end{table}	
We also present the estimation results for jointly modeling LH peak value and TTP in Table~\ref{t:peakv_Joint}. The conclusions of covariate effects were generally consistent with that in Table~\ref{t:kpeak_Joint}, except that obesity seems to have protective effect on TTP and heavy smoking women also seem to have significantly longer TTP than non-smoking women.
\begin{table}
	\caption{Joint model estimation results on LH peak value.}
	\label{t:peakv_Joint}
	\centering
	\fbox{
		\begin{tabular}{rrrr}
			Parameter& Estimate ($95\%$CI) & Parameter& Estimate ($95\%$CI) \\\midrule 
			$\beta$: Age*& $-$0.03 ($-$1.45, 1.38)&$\gamma$: avg.age*& $-$25.41 ($-$39.73, $-$11.09) \\
			$\beta$: Age*$^2$& $-$0.01 ($-$0.47, 0.46)&$\gamma$: dif.age*& $-$5.31 ($-$8.68, $-$1.93) \\
			$\beta$: Overweight& $-$0.05 ($-$0.17, 0.06)&$\gamma$: Overweight& 0.12 ($-$2.81, 3.05) \\
			$\beta$: Obese& $-$0.25 ($-$0.39, $-$0.12)&$\gamma$: Obese& 7.89 (2.47, 13.31) \\
			$\beta$: Smoke.m& 0.06 ($-$0.06, 0.18)&$\gamma$: Smoke.m& $-$7.82 ($-$12.94, $-$2.69)  \\
			$\beta$: Smoke.h& 0.21 ($-$0.06, 0.48)&$\gamma$: Smoke.h& $-$8.86 ($-$17.29, $-$0.44)  \\
			$\beta$: Alp*& $-$0.18 ($-$0.35, $-$0.01)&$\gamma$: Alp*& 2.50 ($-$2.26, 7.27) \\
			$\beta$: Alp*$^2$& 0.04 ($-$0.01, 0.09)&$\gamma$: Alp*$^2$ & $-$0.41 ($-$1.75, 0.92)  \\
			$\beta$: Parity& 0.02 ($-$0.07, 0.12)&$\gamma$: Parity& 6.07 (1.98, 10.16) \\
			$\rho_1$& $-$11.50 ($-$18.68, $-$4.33)& $\rho_2$& 1.34 ($-$0.42, 3.10) \\
			$\rho_3$& 7.97 (4.62, 11.32)& $\rho_4$& 10.35 (5.73, 14.97) \\
			$\rho_5$& 13.48 (7.06, 19.89)& $\rho_6$& 13.81 (6.97, 20.65) \\
			$\rho_7$& $-$5.05 ($-$5.05, $-$5.05)& $\psi_\mu$& 24.63 (9.67, 39.59) \\
			$\sigma_1$& 0.24 (0.20, 0.29)& $\sigma_2$& 25.30 (15.01, 42.65)  \\
			$\sigma$& 0.43 (0.40, 0.45)& $\zeta$& $-$0.22 ($-$0.29, $-$0.14) \\
	\end{tabular}}
\end{table}	
The estimation results for joint modeling of average curvature of LH profile within fertile window and TTP are shown in Table~\ref{t:kavgft_Joint} The overall trend of covariate effects is consistent with that in Table~\ref{t:kpeak_Joint} for curvature at LH peak, except that overweight women seem to have significant longer TTP than underweight/normal weight women, women with higher stress level (alpha amylase) seem to have longer TTP, while the effect of smoking on TTP does not seem to be significant.
\begin{table}
	\caption{Joint model estimation results on average curvature of LH profile within fertile window.}
	\label{t:kavgft_Joint}
	\centering
	\fbox{
		\begin{tabular}{rrrr}
			Parameter& Estimate ($95\%$CI) & Parameter& Estimate ($95\%$CI) \\\midrule 
			$\beta$: Age*& $-$0.27 ($-$2.32, 1.78)&$\gamma$: avg.age*& $-$23.27 ($-$38.97, $-$7.58) \\
			$\beta$: Age*$^2$& 0.10 ($-$0.57, 0.77)&$\gamma$: dif.age*& $-$8.61 ($-$14.08, $-$3.13) \\
			$\beta$: Overweight& $-$0.07 ($-$0.24, 0.09)&$\gamma$: Overweight& $-$8.04 ($-$13.80, $-$2.28) \\
			$\beta$: Obese& $-$0.08 ($-$0.28, 0.12)&$\gamma$: Obese& 2.76 ($-$2.05, 7.57) \\
			$\beta$: Smoke.m& 0.06 ($-$0.12, 0.24)&$\gamma$: Smoke.m& $-$3.98 ($-$8.49, 0.53)  \\
			$\beta$: Smoke.h& 0.07 ($-$0.34, 0.49)&$\gamma$: Smoke.h& 2.93 ($-$7.44, 13.30)  \\
			$\beta$: Alp*& $-$0.18 ($-$0.44, 0.08)&$\gamma$: Alp*& $-$14.97 ($-$25.34, $-$4.59) \\
			$\beta$: Alp*$^2$& 0.04 ($-$0.04, 0.12)&$\gamma$: Alp*$^2$ & 5.45 (1.89, 9.01)  \\
			$\beta$: Parity& 0.02 ($-$0.12, 0.17)&$\gamma$: Parity& 20.65 (9.02, 32.29) \\
			$\rho_1$& $-$27.04 ($-$44.87, $-$9.21)& $\rho_2$& $-$5.70 ($-$12.00, 0.60) \\
			$\rho_3$& 8.28 (5.64, 10.92)& $\rho_4$& 13.38 (8.07, 18.69) \\
			$\rho_5$& 22.83 (12.21, 33.45)& $\rho_6$& 23.56 (11.93, 35.19) \\
			$\rho_7$& $-$0.02 ($-$0.02, $-$0.02)& $\psi_\mu$& 18.53 (4.67, 32.38) \\
			$\sigma_1$& 0.36 (0.30, 0.44)& $\sigma_2$& 43.56 (24.77, 76.59)  \\
			$\sigma$& 0.60 (0.56, 0.64)& $\zeta$& $-$0.17 ($-$0.24, $-$0.10) \\
	\end{tabular}}
\end{table}	

\begin{figure}
	\centering
	\includegraphics[width=.45\textwidth]{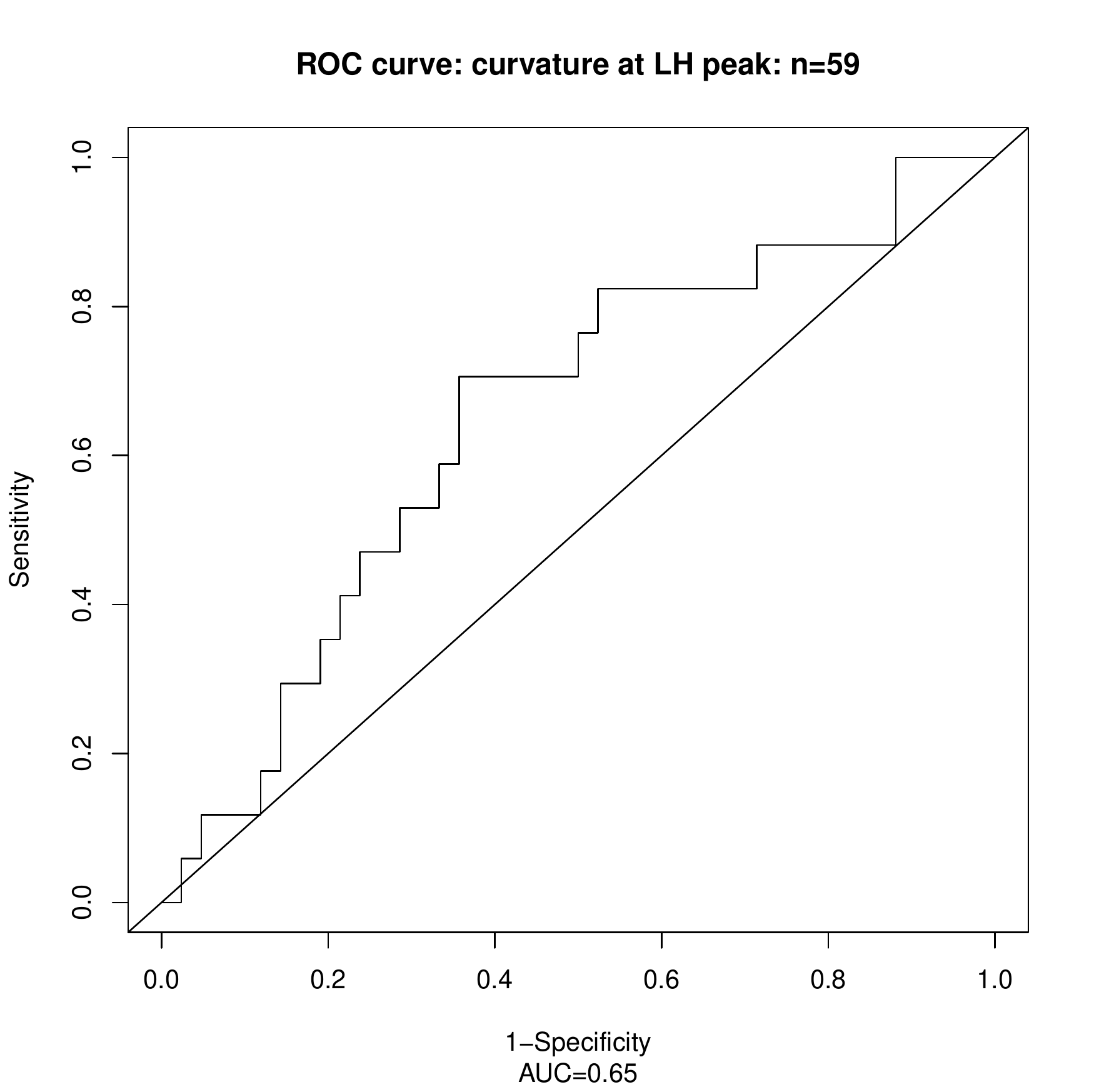}
	\includegraphics[width=0.45\textwidth]{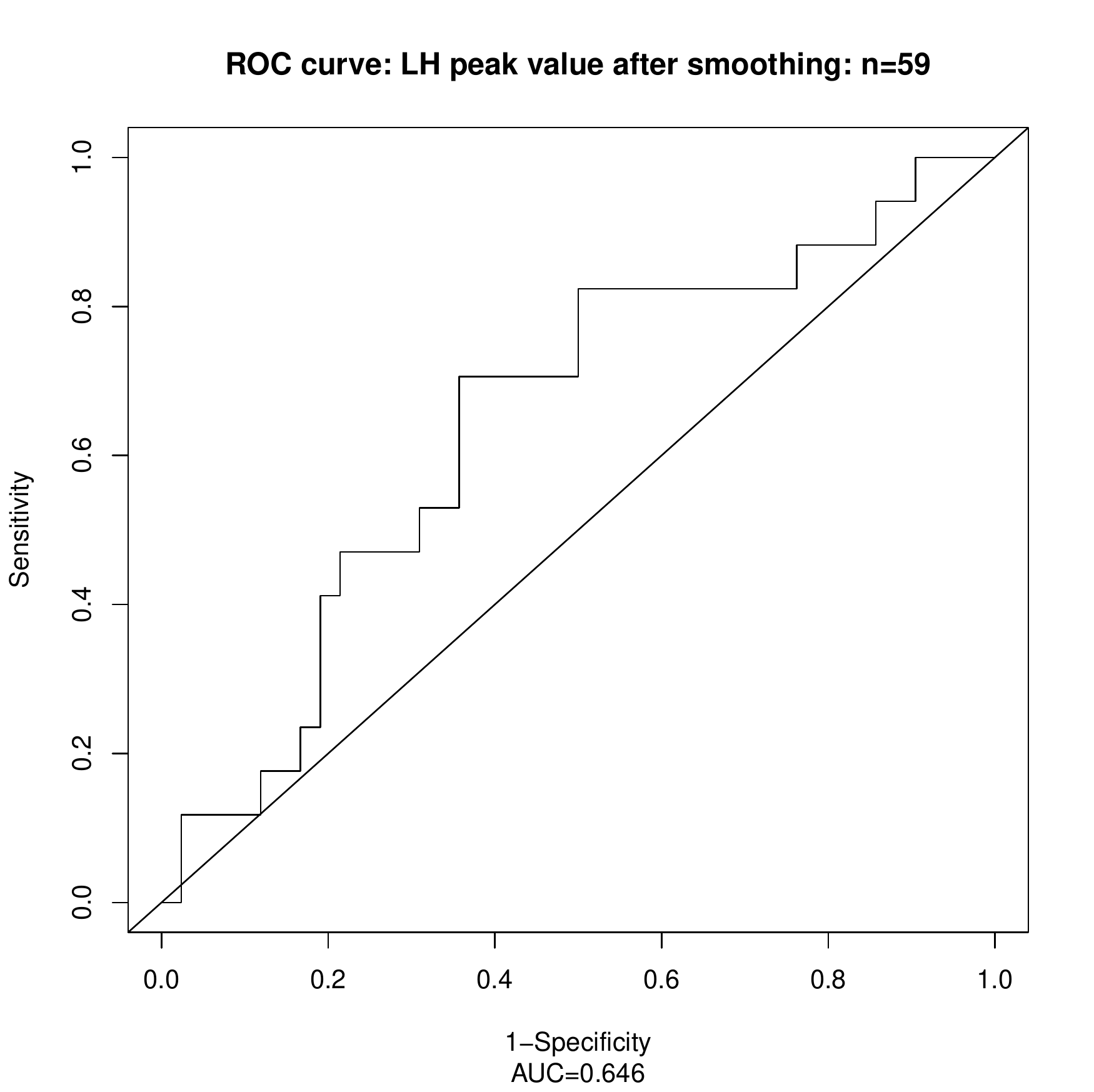}
	\\
	\includegraphics[width=0.6\textwidth]{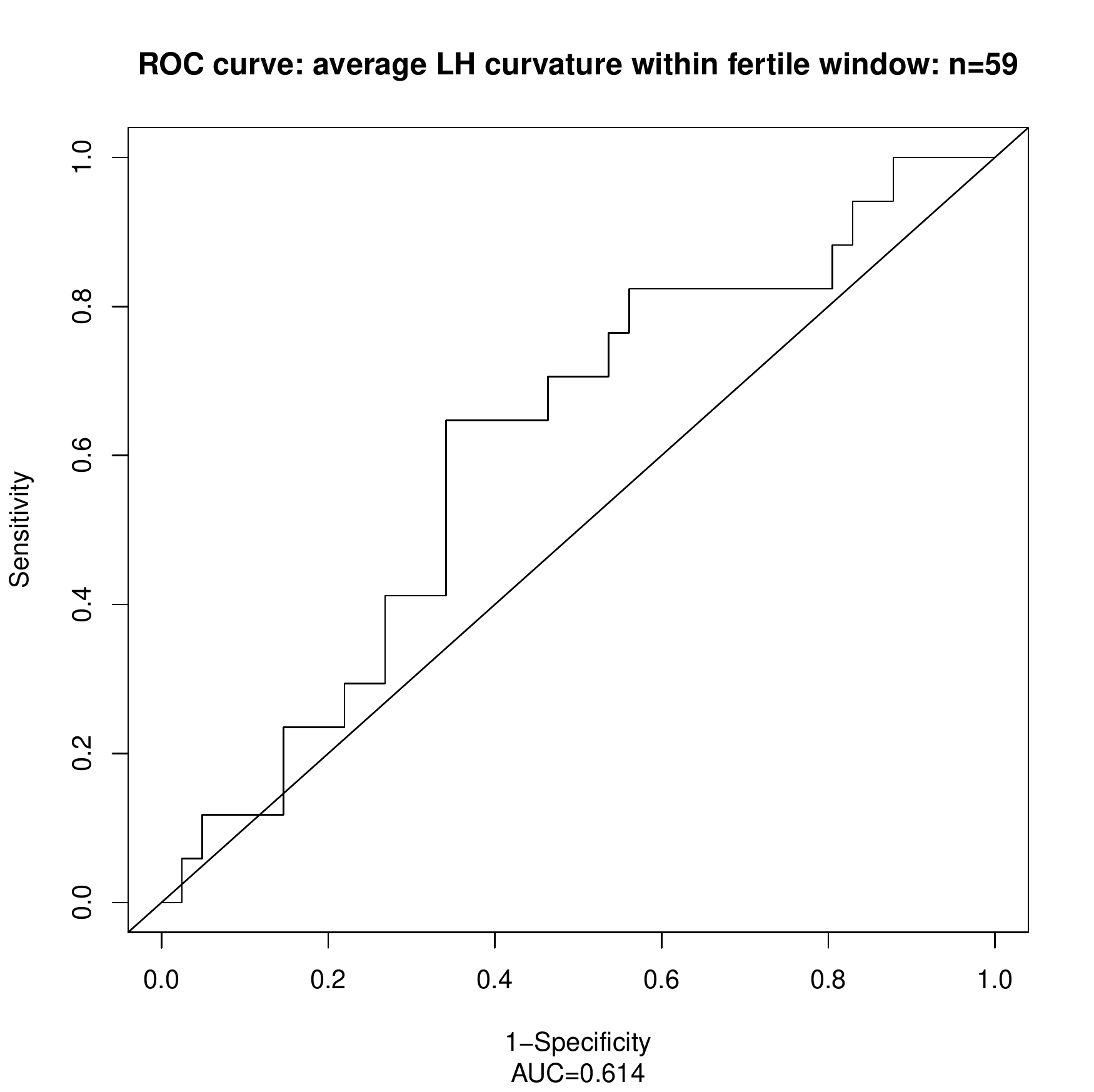}
	\caption{ROC curves for classifying $I(T_i>6)$ by $\hat{\pi}_i(6 \mid 1)$, for models with: (a) curvature at LH peak; (b) LH peak value; (c) average LH curvature within fertile window.}
	\label{fig:ROC}
\end{figure}

Now that we have fitted the joint models using the training dataset, we are interested in predicting the probability of subfertility (i.e., TTP $>6$ cycles) given survival past one cycle. That is, for all women in the prediction set with $T_i>1$, we wish to classify $I(T_i>6)$ by the conditional survival probability $\pi_i(6 \mid 1)$. To measure the classification rate, we empirically estimated the sensitivity: $P(\hat{\pi}_i(6 \mid 1)>c \mid T_i>6)$, and specificity $P(\hat{\pi}_i(6 \mid 1)\leq c \mid T_i\leq 6)$, where $\hat{\pi}_i(6 \mid 1)$ is the mean of the Monte Carlo sample $\{\pi_i^{(l)}, l=1,\dots,L\}$ obtained following the steps in Section~\ref{s:estimation}. Of the 79 women available for prediction, 20 were censored between 1 and 6 cycles and removed from the prediction analysis. The classification measures using ROC for the models with curvature at LH peak, LH peak value and average curvature of LH profile within fertile window, are displayed in Figure~\ref{fig:ROC}(a),(b),(c), respectively, for all $c\in[0,1]$. The AUC is 0.650 for the model which includes the curvature at LH peak, 0.646 for the model which includes the LH peak value and 0.614 for the model which includes the average curvature of LH profile within fertile window. This indicates that the prediction ability of the three models are moderately good.

As mentioned in Section~\ref{s:estimation}, one could use the fitted joint model and hormonal measurement history up to cycle 1, to predict the conditional survival probability $\pi_i(j \mid 1)$, for any $j>1$ if that is of interest. The data used in the above analysis are available from the corresponding author upon reasonable request.

\section{Discussion}
\label{s:discuss}

We have proposed a joint modeling approach to assess the association between cycle-level geometric features of a woman's hormonal profile and her fecundity measured by TTP. A likelihood based approach with the use of Gaussian quadrature approximation was proposed for estimation of the unknown parameters. Simulation studies have demonstrated that the proposed estimation approach works reasonable well in the situation similar to the Oxford data, with reasonably small bias and coverage probabilities close to the nominal level. With the estimates from the joint models, we also derived the approach to predict individual characteristics of TTP given a set of longitudinal measurements up to a certain cycle. The prediction of the probability that a woman was subinfertile given her past one menstrual cycle behavior without getting pregnant, was moderately accurate for model with any one of the three geometric features of LH profile, especially with curvature at hormonal profile peak (AUC around 0.65).

The analysis of Oxford data found that couple average age and difference between female and male age are signifcantly associated with TTP in the sense that older couples and couples with larger female minus male age difference have significantly longer TTP. The association between BMI and TTP seem to be marginal with moderate evidence that overweight women have a slower rate of pregnancy while obese women have a faster rate of pregnancy. Multiparous women were found to have significantly shorter TTP than nulliparous women. Furthermore, we found that women with sharper LH peaks, higher LH peaks and overall more curved LH profiles within fertile window tend to have significantly shorter TTP.  

We focused on cycle-varying geometric features of the hormonal profile, motivated by biology as well as due to the availability of hormonal data only around ovulation. In the scenario where hormonal profile is available throughout the menstrual cycle, a more general approach of considerable interest is to look at the full hormonal profile and assess patterns in the framework of 
joint modeling of functional data and time to event. Such approach have been studied by Li et al.(2017,2019,2022)\citep{li2017functional,li2019bayesian,li2022joint} and may be extended in this context of cyclical hormonal profile and TTP to shed light on meaningful patterns on TTP, as well as its predictive ability on infertility. Another interesting approach along this line is the recent development of the functional model proposed for the cyclical longitudinal process \citep{ji2020semiparametric}, which can be extended in the context of joint modeling.

Finally, our approach though motivated by reproductive epidemiology may be applicable to examples arising in other disciplines. For example, one may be interested in understanding the association between the high blood pressure and risk for heart attack and not just on their average blood pressure measurements. In conclusion, this paper studies a novel model for joint models of longitudinal and survival data where one is interested in the longitudinally varying geometric features with survival data.

\backmatter


\bmhead{Acknowledgments}

This research was supported in part by the Intramural Research Program of the National Institutes of Health, \emph{Eunice Kennedy Shriver} National Institute of Child Health and Human Development. The authors would like to acknowledge that this study utilized the high-performance computational capabilities of the Biowulf Linux cluster at the National Institutes of Health, Bethesda, MD (\url{http://biowulf.nih.gov}).
The research of Animikh Biswas was supported in part by the NSF grant DMS 1517027.

\noindent
{\it Conflict of Interest}: None declared.
\vspace*{-8pt}

\abhi{
	\newpage
\begin{appendices}
\section{ Simulation results: p($A_{ij}$)=0.90 and =0.85}\label{AppTables}	
\vspace{-0.9cm}
	\begin{table}[ht!]  
		\centering
	 \begin{minipage}{\textwidth}
		\caption{Simulation results corresponding to varying sample sizes with $p(A_{ij})=0.90$}    \label{tab2}  
		\small
\begin{tabular}{@{\extracolsep{\fill}}l@{\extracolsep{\fill}}c@{\extracolsep{\fill}}|ccl|@{\extracolsep{\fill}}ccc|@{\extracolsep{\fill}}ccc@{\extracolsep{\fill}}}
	\toprule
	\multicolumn{2}{@{\extracolsep{\fill}}c@{\extracolsep{\fill}}}{Param.} & \multicolumn{3}{c@{\extracolsep{\fill}}}{n=300, $n_{tr}$=200} & \multicolumn{3}{c}{n=400, $n_{tr}$=267} & \multicolumn{3}{c}{n=500, $n_{tr}$=334} 
	\\\hline
			Symb. & True & Bias  & SD & CP & Bias  & SD & CP & Bias  & SD & CP \\\hline
			$\beta_1$ & 4 & 0.006 & 0.125 & 0.953 & -0.003 & 0.114 & 0.946 & -0.006 & 0.11 & 0.938\\
			$\beta_2$ & -0.5 & -0.005 & 0.049 & 0.947 & 0 & 0.044 & 0.95 & 0.001 & 0.043 & 0.945\\
			$\gamma_1$ & -27 & -0.099 & 1.654 & 0.968 & -0.119 & 1.559 & 0.972 & -0.084 & 1.52 & 0.945\\
			$\rho_1$ & -6.5 & 0.538 & 2.313 & 0.965 & 0.503 & 2.15 & 0.963 & 0.493 & 2.009 & 0.952\\
			$\rho_2$ & 10 & 0.608 & 2.08 & 0.96 & 0.433 & 1.933 & 0.97 & 0.538 & 1.743 & 0.971\\
			$\rho_3$ & 17 & 0.589 & 2.044 & 0.941 & 0.431 & 1.78 & 0.925 & 0.478 & 1.708 & 0.935\\
			$\rho_4$ & 21 & 0.626 & 1.955 & 0.962 & 0.498 & 1.773 & 0.961 & 0.465 & 1.727 & 0.956\\
			$\rho_5$ & 24 & 0.701 & 2.05 & 0.96 & 0.599 & 1.814 & 0.97 & 0.546 & 1.77 & 0.954\\
			$\rho_6$ & 25 & 0.349 & 2.318 & 0.975 & 0.443 & 2.017 & 0.975 & 0.427 & 1.95 & 0.957\\
			$\psi_{\mu}$ & 18 & -0.058 & 1.557 & 0.971 & -0.014 & 1.503 & 0.966 & -0.055 & 1.418 & 0.944\\
			$\sigma_1$ & 0.3 & 0.009 & 0.076 & 0.909 & 0.004 & 0.062 & 0.923 & 0.009 & 0.06 & 0.917\\
			$\sigma_2$ & 3 & 0.771 & 3.053 & 0.955 & 0.562 & 2.669 & 0.962 & 0.5 & 1.759 & 0.953\\
			$\sigma$ & 0.9 & -0.004 & 0.029 & 0.946 & -0.003 & 0.028 & 0.944 & -0.002 & 0.025 & 0.941\\
			$\zeta$ & -0.2 & -0.01 & 0.463 & 0.902 & 0.007 & 0.419 & 0.918 & -0.009 & 0.383 & 0.913\\
			\botrule
		\end{tabular}
	\end{minipage}
	\end{table}
\vspace{-01cm}
\begin{table}[ht!]  
	\centering
	\caption{Simulation results corresponding to varying sample sizes with  $p(A_{ij})=0.85$}    \label{tab3}  
	\small
	\begin{tabular}{@{\extracolsep{\fill}}l@{\extracolsep{\fill}}c@{\extracolsep{\fill}}|ccl|@{\extracolsep{\fill}}ccc|@{\extracolsep{\fill}}ccc@{\extracolsep{\fill}}}
		\toprule
		\multicolumn{2}{@{\extracolsep{\fill}}c@{\extracolsep{\fill}}}{Param.} & \multicolumn{3}{c@{\extracolsep{\fill}}}{n=300, $n_{tr}$=200} & \multicolumn{3}{c}{n=400, $n_{tr}$=267} & \multicolumn{3}{c}{n=500, $n_{tr}$=334} 
		\\\hline
		Symb. & True & Bias  & SD & CP & Bias  & SD & CP & Bias  & SD & CP \\\hline
		$\beta_1$ & 4 & 0.009 & 0.129 & 0.94 & 0.004 & 0.109 & 0.949 & -0.007 & 0.1 & 0.941\\
		$\beta_2$ & -0.5 & -0.006 & 0.051 & 0.947 & -0.003 & 0.043 & 0.953 & 0.001 & 0.039 & 0.938\\
		$\gamma_1$ & -27 & -0.148 & 1.939 & 0.976 & -0.083 & 1.557 & 0.973 & -0.119 & 1.445 & 0.953\\
		$\rho_1$ & -6.5 & 0.437 & 2.485 & 0.977 & 0.558 & 2.089 & 0.969 & 0.433 & 2.087 & 0.959\\
		$\rho_2$ & 10 & 0.493 & 2.445 & 0.962 & 0.511 & 1.826 & 0.973 & 0.496 & 1.775 & 0.959\\
		$\rho_3$ & 17 & 0.52 & 2.254 & 0.94 & 0.575 & 1.736 & 0.947 & 0.493 & 1.685 & 0.942\\
		$\rho_4$ & 21 & 0.614 & 2.128 & 0.965 & 0.567 & 1.783 & 0.965 & 0.543 & 1.787 & 0.929\\
		$\rho_5$ & 24 & 0.693 & 2.205 & 0.976 & 0.678 & 1.983 & 0.958 & 0.566 & 1.833 & 0.953\\
		$\rho_6$ & 25 & 0.501 & 2.53 & 0.972 & 0.497 & 2.164 & 0.97 & 0.367 & 1.77 & 0.964\\
		$\psi_{\mu}$ & 18 & -0.021 & 1.87 & 0.97 & -0.053 & 1.485 & 0.975 & -0.009 & 1.403 & 0.96\\
		$\sigma_1$ & 0.3 & 0.008 & 0.073 & 0.917 & 0.008 & 0.065 & 0.918 & 0.008 & 0.061 & 0.918\\
		$\sigma_2$ & 3 & 0.67 & 2.32 & 0.949 & 0.713 & 3.117 & 0.962 & 0.512 & 2.313 & 0.949\\
		$\sigma$ & 0.9 & -0.003 & 0.031 & 0.941 & -0.002 & 0.028 & 0.931 & -0.002 & 0.024 & 0.946\\
		$\zeta$ & -0.2 & -0.002 & 0.485 & 0.894 & -0.011 & 0.414 & 0.929 & -0.007 & 0.382 & 0.92\\
		\botrule
	\end{tabular}
\end{table}
\vspace{-0.9cm}
\section{Longitudinal data at each cycle}\label{Long}
 In the real data applications, the geometric features were calculated from the (smoothed) hormonal profiles obtained from the data while in the simulation study, these were generated directly from a linear model. We show below that under certain specific simulation schemes, they are equivalent.

We first generate $Y_{ij}$, and $\hat{Y}_{ij}$ once $Z_{ij}$ and $b_{Y,i}$ are simulated as given below (See (\ref{hormone_model}) and Section \ref{s:simulation}).
\begin{equation*} 
	Y_{ij}=\tilde{Y}_{ij}+\epsilon_{ij}, \tilde{Y}_{ij}=Z_{ij}'\bss{\beta} + b_{Y,i}, \label{generated}
\end{equation*}
where $\epsilon_{ij} \sim \textrm{N}(0,\sigma^2)$. To generate a true hormonal profile and the observed hormonal profile we consider a class of functions that are symmetric around 0 with a peak at 0.  We further assume they both belong in this class of functions. For illustration purposes, let us consider the class  $\mathcal{G}_{\lambda}$, given by the scaled truncated centered Gaussian curves parametrized by the precision parameter. 
\begin{align*}
	\mathcal{G}_{\lambda} =\{h:  h(t; \lambda)= \lambda \exp(-\lambda t^2), \lambda \in \mathbb{R}^{+}, t \in [-T_{ij}, T_{ij}] \}
\end{align*}
Let $\mathcal{F}$ be the functional of interest corresponding to a given geometric feature. Define the value of  geometric feature, $g(\lambda)$, as follows,
\begin{equation}
	g(\lambda)= \mathcal{F}(h(.; \lambda)), h \in \mathcal{G}_{\lambda}
	\label{gfunction}
\end{equation}
$\mathcal{F}:$ value at $t=0$ (at peak) $\Rightarrow g(\lambda)= \mathcal{F}(h) = h(0)= \lambda$. Similarly, $\mathcal{F}:$ curvature at peak   $g(\lambda)= 2\lambda^2$. A similar formula for $g(\lambda)$ can be worked out for average curvature.
Thus, when the value of peak is modeled, one can generate a true hormonal profile curve and an observed hormonal curve, and compute them at a given time point $t \in [-T_{ij}, T_{ij}] $ as follows,
\begin{equation*}
  \tilde{H}_{ij,t} = h_1(t)  =h(t;\tilde{Y}_{ij}); \, H_{ij,t}  =h_2(t) =h(t ;Y_{ij}),  \,\, h_1, h_2 \in  \mathcal{G_\lambda}  
\end{equation*}
 Thus, for any general  $\mathcal{F}$ representing a geometric feature, $h_1$ and $h_2$ can be chosen with appropriate choices of $\lambda$ as follows,
\begin{equation}
\tilde{H}_{ij,t} = h_1(t) =h(t; g^{-1}(\tilde{Y}_{ij})); \, H_{ij,t}  =h_2(t) =h(t ;g^{-1}(Y_{ij}))  \,\, \label{ginvfunction} 
\end{equation}
In particular, for curvature at peak, $h_1(t)=h(t, \sqrt{\tilde{Y}_{ij}/2} )$, and $ h_2(t)=h(t, \sqrt{Y_{ij}/2} )$. Let $\eta_{ij,t}= H_{ij,t}- \tilde{H}_{ij,t}$ be the nested random error observed at $t$. Note that  $\eta_{ij,t}$ is simulated implicitly where $\tilde{H}_{ij,t}$'s   constitute unobserved but true hormonal curve for a given individual. 
Thus, if a sufficiently large number of $t$'s are observed or a good smoothing process is applied,  the $H$ function can be approximated well, leading to the recovery of observed $Y_{ij}$ values upon the application of $\mathcal{F}$ on $H$. Since the joint model summarizes data at the cycles, generating $H$ becomes redundant. 

Note that the simulation scheme mentioned above can be implemented to a much larger class of function than $\mathcal{G}_\lambda $, which is chosen for illustration purposes only, to model hormonal profiles. Assuming that the hormonal curve is twice continuously differentiable with peak at $t=0$, we must have $\tilde{H}^{'}_{ij}(t)=0$. Thus, the value at the peak is $\tilde{H}_{ij}(0)$ and the curvature at the peak is $\frac{\vert\tilde{H}^{''}_{ij}(0)\vert}{(1+\tilde{H}^{'2}_{ij}(0))^{3/2}}=\vert\tilde{H}^{''}_{ij}(0)\vert$. Accordingly, let $\mathcal{C}_f$ denote such a class of functions. Then, a larger two-parameter scale family of functions, $\mathcal{G}_{\lambda_1, \lambda_2}$, based on $\mathcal{C}_f$ can be used to model hormonal profile curves as given below.

\begin{align*} \label{scalefamily}
	{\mathcal{C}_f} &= \{\Psi:{\mathbb R} \rightarrow {\mathbb R}, \Psi''\  \mbox{is continuous}, \Psi \ \mbox{has a unique maximum at }\ t=0\} \\
	{\mathcal G}_{\lambda_1, \lambda_2} &=
	\left\{h(t;\lambda_1,\lambda_2)= \lambda_1 f\left(\sqrt{\frac{\lambda_2}{\lambda_1}}\ t\right), t \in [-T_{ij},T_{ij}], f \in \mathcal{C}_f \right\},
\end{align*}
Note $\mathcal{G}_\lambda \subset {\mathcal G}_{\lambda_1, \lambda_2}$  with $\lambda_1=\lambda$, $\lambda_2=\lambda^2$, $f= e^{-s^2}$. Similar to (\ref{gfunction}) one can define $g(\lambda_1,\lambda_2)$ [=$c_1\lambda_1, c_1= f(0)$ for peak; = $ c_2\lambda_2, c_2= \vert f^{''}(0)\vert$  for curvature at peak]. Similar to (\ref{ginvfunction}), one can simulate true and observed hormonal curves as an intermediate step by choosing $h_1, h_2$ with appropriate $\lambda_1$ and  $\lambda_2$ values. These choices  of $h_1$ and $h_2$ often assume either $Y_{ij} , \tilde{Y}_{ij}>0$ always, which in general would not be true, but can be made to practically hold true by controlling $\beta$'s and choosing a small $\sigma$ relative to the mean. Alternatively, one can apply an appropriate monotonic  function (such as logarithm to peak value) before applying the linear model during data preprocessing.
\end{appendices}
}
\bibliography{Reference}
\end{document}